\documentclass[11pt]{article} 
\usepackage[utf8]{inputenc}
\usepackage{multirow}
\usepackage{amsfonts}
\usepackage{amsmath}
\usepackage{amssymb}
\usepackage{amsthm}
\usepackage{caption}
\usepackage[dvipsnames]{xcolor}
    \definecolor{darkgreen}{rgb}{0,0.5,0}
    \definecolor{darkblue}{rgb}{0,0,0.6}
    \definecolor{purple}{rgb}{0.4,.2,0.7}
\usepackage[margin = 2.5cm]{geometry}
\usepackage{graphicx}
\usepackage[hyperfootnotes = false, colorlinks = true, linkcolor = darkblue, citecolor = purple]{hyperref}
\usepackage{subcaption}
\usepackage{ytableau}

\usepackage{csquotes}

\usepackage{tikz}
\usetikzlibrary{decorations.pathmorphing}
\usetikzlibrary{decorations.markings}
\definecolor{mathred}{RGB}{180,44,37}
\definecolor{mathblue}{RGB}{39,94,190}
\tikzset{>=latex} 
\tikzset{ photon/.style={decorate, decoration={snake}, draw=black}}

\usepackage{comment}

\usepackage{physics}

\newcommand{\be}{\begin{equation}}
\newcommand{\ee}{\end{equation}}
\newcommand{\bea}{\begin{eqnarray}}
\newcommand{\eea}{\end{eqnarray}}

\begin{document}

\thispagestyle{empty}
\begin{center}
    ~\vspace{5mm}

  \vskip 2cm 
  
   {\LARGE \bf 
       Higher-point correlators in the BFSS matrix model
   }

   \vspace{0.5in}
     
   {\bf Anna Biggs$^1$ and Aidan Herderschee$^2$
   }

    \vspace{0.5in}

 $^1$
  Department of Physics, Princeton University, \\
  Jadwin Hall, Washington Road, Princeton, NJ 08544, USA
   \\
   ~
   \\
  $^2$
  School of Natural Sciences, Institute for Advanced Study, \\
  1 Einstein Drive, Princeton, NJ 08540, USA

    \vspace{0.3in}

    E-mail: \texttt{abiggs@princeton.edu}, \texttt{aidanh@ias.edu}

    \vspace{0.5in}

\end{center}

\vspace{0.5in}

\begin{abstract}

We investigate higher-point correlators in the BFSS matrix model, a non-conformal theory dual to type IIA string theory, using Witten diagrams and techniques from the amplitudes program. While the Witten diagrams are more complex than those relevant for conformal holographic theories, we find that the added complexity is minimal. As an illustration, we compute the leading three-point Witten diagram, evaluating it in a squeezed limit using the method of regions. Our results provide new targets for future Monte-Carlo and quantum simulations of the BFSS matrix model.

\end{abstract}

\vspace{1in}

\pagebreak

\setcounter{tocdepth}{3}
{\hypersetup{linkcolor=black}\tableofcontents}

\section{Introduction}\label{sec:introduction}

The BFSS matrix model is conjectured to be dual to the near-horizon geometry of D0-branes in IIA string theory \cite{deWit:1988wri,Banks:1996vh,Susskind:1997cw,Seiberg:1997ad,Sen:1997we,Itzhaki:1998dd,Polchinski:1999br}. It exhibits holographic duality in the spirit of AdS/CFT \cite{Maldacena:1997re,Witten:1998qj,Gubser:1998bc}. Unlike many other holographic models, it is a purely quantum mechanical system rather than a quantum field theory. Moreover, in contrast to other quantum mechanical models exhibiting holographic behavior, such as SYK \cite{Polchinski:2016xgd,Maldacena:2016hyu,Kitaev:2017awl}, the BFSS model has a dual description in terms of a local gravity theory governed by the Einstein-Hilbert action. In other words, the higher-spin bulk states, corresponding to excited string states, are parametrically heavy. Consequently, the BFSS matrix model serves as an ideal setting for precision studies of quantum gravity using Monte-Carlo and quantum simulations. Indeed, Monte-Carlo calculations have already confirmed several nontrivial predictions of the holographic dual \cite{Catterall:2007fp, Hanada:2007ti, Anagnostopoulos:2007fw, Catterall:2008yz, Hanada:2008ez, Hanada:2008gy,  Catterall:2009xn,  Hanada:2009ne, Hanada:2011fq, Filev:2015hia, Berkowitz:2016jlq,Berkowitz:2018qhn,Bergner:2021goh,Pateloudis:2022ijr}.

Computations of higher point correlators via the holographic duality have already been done in many other holographic models, such as $\mathcal{N}=4$ super-Yang Mills, see Refs. \cite{DHoker:1998bqu,DHoker:1998ecp,Liu:1998th,DHoker:1999bve,DHoker:1999kzh,DHoker:1999mqo} for early papers. In contrast, the higher point correlators of the BFSS matrix model are unexplored. In this letter, we compute higher point correlation functions  
\begin{equation}
\langle \mathcal{O}(t_{1})\mathcal{O}(t_{2})\ldots \rangle
\end{equation}
in the regime 
\begin{equation}\label{con1}
1<<N, \quad N^{-10/21}<<T\lambda^{-1/3}<<1
\end{equation}
using the holographic duality, where $T$ is the temperature, $\lambda$ is the 't Hooft coupling and $N$ is the size of the matrices. In this regime, correlators can be computed using Witten diagrams in ten-dimensional IIA supergravity.\footnote{This is distinct from the eleven-dimensional supergravity regime that describes dynamics at parametrically low energy at large-$N$. This regime is traditionally probed by scattering amplitudes \cite{Miller:2022fvc,Tropper:2023fjr,Herderschee:2023pza,Herderschee:2023bnc}, not finite-time correlators.} We additionally impose that the time-scale of the operator insertions obeys 
\begin{equation}\label{eq:holographicrenorm}
\forall i\neq j:\quad T^{-1}>>|t_{i}-t_{j}|>>\lambda^{-1/3} 
\end{equation}
to simplify calculations. We will restrict to operators separated in Euclidean time because it is difficult to study real-time dynamics using Monte-Carlo due to the sign problem.

\begin{figure}[t]
\centering
\includegraphics[width=8cm]{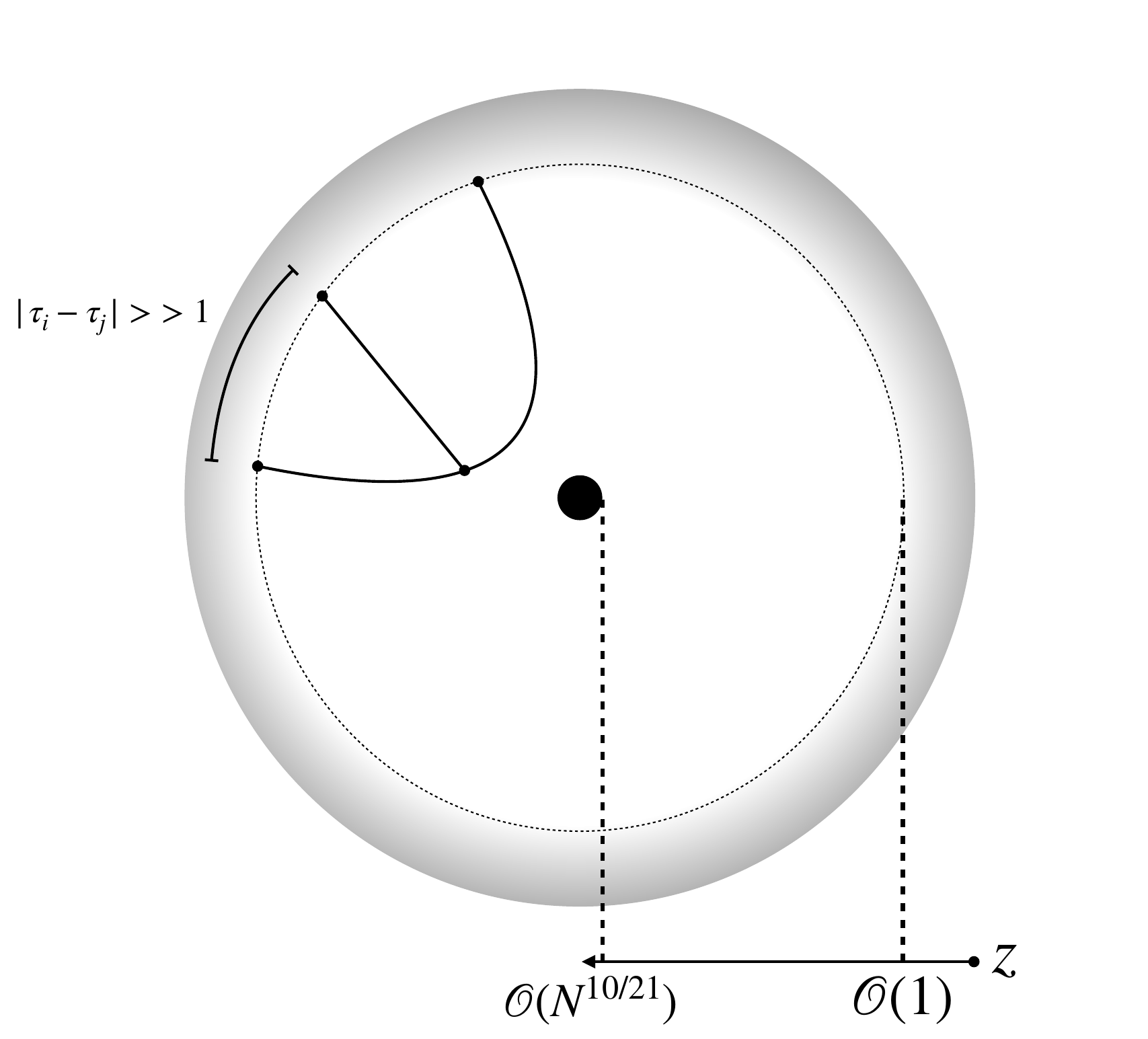}
\caption{A sketch of the gravity dual and the correlators we are studying in $z,\tau$ coordinates given by Eq. (\ref{metricintro}) in string units. The near boundary region, $z<<1$, is highly curved and the gravity approximation cannot be trusted. The boundary in the deep interior corresponds to when the type IIA supergravity approximation breaks down. We insert operators at the boundary of the supergravity region as prescribed by holographic renormalization at time scales that are parametrically large in string units.}\label{fig:gravdual}
\end{figure}
The main challenge in studying these correlators is that the BFSS matrix model lacks conformal or approximate conformal symmetry. The bulk dual of more traditional holographic models is the product of Anti-de Sitter (AdS) with some compact space. The fact that AdS is a maximally symmetric spacetime maps to a conformal symmetry of the boundary theory. Conformal symmetry is sufficient to uniquely fix the kinematic dependence of the two and three-point functions. At low temperatures, the holographic dual of BFSS in the relevant regime is not $\textrm{AdS}_{2}\times \textrm{S}_{8}$, but instead a Weyl re-scaling that breaks the AdS isometry group:
\begin{equation}\label{metricintro}
ds^{2}\propto z^{3/5}\left [ \left ( \frac{2}{5}\right )^{2}\left (\frac{d\tau^{2}+dz^{2}}{z^{2}}\right )+d\Omega_{8}^{2}  \right ]  \ ,
\end{equation}
where $\tau\propto \lambda^{1/3}t$. There is also a non-trivial dilaton background. A sketch of the gravity dual and the regime in which we are computing correlators is shown in Fig.~\ref{fig:gravdual}. Due to the lack of conformal symmetry, the position dependence of higher point correlators is much less constrained. In addition to this technical difficulty, there are some conceptual subtleties. In particular, the supergravity approximation of the bulk is not valid at all energy scales \cite{Itzhaki:1998dd}. For example, the dynamics of the near boundary region is not well described by supergravity.

However, none of these issues are insurmountable and standard technology, such as Witten diagrams \cite{Witten:1998qj,Freedman:1998tz,DHoker:1999mqo}, can be applied to study correlators in the regime given by Eqs. (\ref{con1}) and (\ref{eq:holographicrenorm}). Witten diagrams provide an algorithm for computing the boundary correlators in the supergravity approximation of the bulk theory by integrating over bulk-to-boundary and bulk-to-bulk propagators in a manner similar to Feynman diagrams. We show that the bulk-to-boundary and bulk-to-bulk propagators relevant for the metric in Eq. (\ref{metricintro}) are almost exactly the same as those of AdS${}_{2}$. The only difference is that each bulk vertex includes some additional factors of $z$ that break conformal symmetry. Therefore, although the loss of conformal symmetry results in more complex integrals, the added complexity is arguably minimal.\footnote{Related phenomena were noted for the Witten diagrams of FLRW spacetimes in Ref. \cite{Arkani-Hamed:2023kig} and for giant gravitons in non-conformal backgrounds in Ref. \cite{Batra:2025ivy}.} 

We compute the leading three-point correlators at strong coupling in terms of a two dimensional Euler integral \cite{aomoto1975equations,gel1986general,Gelfand:1990bua}.\footnote{See Ref. \cite{Matsubara-Heo:2023ylc} for review.} Euler integrals are defined as 
\begin{equation}
\int \left [\prod_{i} du_{i} u_{i}^{s_{i}}\right ]\left [\prod_{j} p_{j}^{-t_{j}}\right ]
\end{equation}
where the $p_{i}$ are Laurent polynomials of the $u_{i}$, and possibly other variables, and $s_{i},t_{i}$ are real numbers. These integrals can be evaluated using numerics. We use the method of regions \cite{Smirnov:1991jn,Beneke:1997zp,Smirnov:1998vk,Smirnov:1999bza,Pak:2010pt,Ananthanarayan:2018tog,Ananthanarayan:2020ptw} to investigate this integral in different limits. In particular, if we impose that
\begin{equation}
\Delta _1+\frac{9}{5}<\Delta _2+\Delta _3\quad  \textrm{and} \quad \tau_{1}>\tau_{2}>\tau_{3} \ ,
\end{equation}
we find an interesting scaling behavior in the squeezed limit:
\begin{equation}\label{scalebehavior}
\textrm{Fixed }(\tau_{1}-\tau_{2}): \ \lim_{\left ( \frac{\tau_{2}-\tau_{3}}{\tau_{1}-\tau_{2}}\right )\rightarrow 0} \langle \mathcal{O}_{\Delta_{1}}(\tau_{1})\mathcal{O}_{\Delta_{2}}(\tau_{2})\mathcal{O}_{\Delta_{3}}(\tau_{3})\rangle \propto \left ( \frac{\tau_{2}-\tau_{3}}{\tau_{1}-\tau_{2}} \right )^{ \Delta _1-\Delta _2- \Delta _3+\frac{9}{5} }
\end{equation}
where the $\Delta_{i}$ correspond to the dimension of the operators in the strong coupling regime.\footnote{We follow the notation in Ref. \cite{Biggs:2023sqw}, not Ref. \cite{Sekino:1999av}. Our $\Delta$ is not the same as Ref. \cite{Sekino:1999av}.} We cross-check our result using an uplift to AdS${}_{2+9/5}$ proposed in Refs. \cite{Kanitscheider:2009as,Biggs:2023sqw} in Section \ref{dilatonuplift}. This scaling behavior is a concrete probe of local bulk dynamics that could be verified via future Monte-Carlo and quantum simulations.

We conclude by showing how these methods can be easily generalized to higher-point and higher-loop contributions using a Schwinger-like formula for the bulk-to-bulk propagator. Although we restrict ourselves to BFSS, the techniques developed in this letter should be applicable to general holographic theories and computations in de Sitter backgrounds.

\paragraph{Note added.} Shortly after the completion of this work, Bobev et al.~\cite{Bobev:2025idz} appeared with results that overlap with several of those presented here. Their work was carried out independently.

\section{Review of the BFSS matrix model}

 In this section, we review the BFSS model, its gravitational dual at finite temperature and the holographic map. The BFSS model is defined by the action 
\begin{equation}\label{bfssaction}
S=\frac{N}{\lambda}\int dt \textrm{Tr}\left [\frac{1}{2}\sum_{i}(D_{t}X^{i})^{2}+\frac{1}{2}\sum_{\alpha}\psi_{\alpha}D_{t}\psi_{\alpha}+\frac{1}{4}\sum_{i,j}[X^{i},X^{j}]^{2}+\sum_{i,\alpha,\beta}\frac{1}{2}i\psi_{\alpha}\gamma_{\alpha\beta}^{i}[\psi_{\beta},X^{i}] \right ]
\end{equation}
where $i,j\in \{1,\ldots,9 \}$ and $\alpha,\beta\in \{1,\ldots,16\}$. The $X^{i}$ are nine flavors of $N\times N$ matrices that correspond to the $N$ D0-branes coupled with open string modes. The $\psi$ are sixteen hermitian $N\times N$ matrices. The $\gamma^{i}_{\alpha\beta}$ are nine dimensional gamma matrices. This model has sixteen linearly realized supercharges and obeys an SO(9) flavor symmetry. Since the model is one-dimensional, the gauge field is not dynamical and amounts to projecting onto the singlet states in the gauge theory. 

\subsection{The gravity dual}\label{gravitydual}
 In this section, we review the holographic dual of BFSS, focusing on the regime of validity of the supergravity approximation of bulk dynamics. The gravity dual is the near horizon limit of the black D0 branes, which in string frame is given by \cite{Horowitz:1991cd,Itzhaki:1998dd}:
\begin{equation}\label{stringframemetrci}
\begin{split}
\frac{ds^{2}}{\alpha'}&=\sqrt{\frac{\lambda d_{0}}{r^{3}}}\left ( -\frac{h(r)r^{5}}{\lambda d_{0}}dt^{2}+\frac{dr^{2}}{h(r)r^{2}}+d\Omega_{8}^{2} \right ) \ , \\
e^{\phi}&=\frac{(2\pi)^{2}}{d_{0}}\frac{1}{N}\left ( \frac{\lambda d_{0}}{r^{3}}\right )^{7/4} \ , \\
A_{t}&=\frac{N}{(2\pi)^{2}}\frac{r^{7}}{\lambda^{2}d_{0}}\ ,  \\
h(r)&=1-\frac{r_{0}^{7}}{r^{7}}, \quad d_{0}=240\pi^{5} \ ,
\end{split}
\end{equation}
where 
\begin{equation}
\frac{1}{T}=\frac{4\pi}{7}\sqrt{\lambda d_{0}}r_{0}^{-5/2} \ .
\end{equation}
The $t$ in Eq. (\ref{stringframemetrci}) should be identified with the $t$ in the BFSS action (\ref{bfssaction}). The background is a Weyl re-scaling of AdS with a charged black brane in the interior. The background has a non-trivial dilaton profile, which reflects how the dual BFSS model is not a conformal theory. 

The validity of the supergravity approximation outside the horizon requires 
\begin{equation}\label{nearhorizon}
N^{-10/21}<<T\lambda^{-1/3}<<1 \ .
\end{equation}
The upper bound comes from imposing that the curvature is not too large at the horizon. The lower bound comes from imposing that the dilaton is not too large at the horizon. Assuming Eq. (\ref{nearhorizon}), the supergravity approximation is trustworthy. In principle, this regime is simultaneously accessible via current Monte-Carlo and holography. However, analyzing finite temperature black branes in AdS is technically difficult even without the Weyl factor. Therefore, we will assume that we are working at low temperatures,
\begin{equation}\label{assumption3}
T\rightarrow 0
\end{equation}
which implies that $r_{0}\rightarrow 0$ and sets $h(r)=1$.\footnote{For example, this assumption would be valid for generic observables if we assumed that the temperature scaling was parametric in $N$:
\begin{equation}
T\lambda^{-1/3}\propto N^{\alpha}, \quad \frac{-10}{21}<\alpha <0 \ .
\end{equation}} This low-temperature assumption drastically simplifies the analysis of the bulk fields. We discuss the validity of this low-temperature assumption in Section \ref{threepointcorrelator}. In particular, if the operator insertions on the boundary are sufficiently localized in time relative to the temperature scale $T$, the low-temperature assumption (\ref{assumption3}) is valid even at finite temperature for the purpose of computing correlators.

Another issue is that the effective curvature of the S${}_8$ sphere, $\sqrt{r^{3}/\lambda d_{0}}$, becomes large in string units when
\begin{equation}\label{nearboundcond}
r \gtrsim r_{b}=(d_{0}\lambda)^{1/3}
\end{equation}
which implies that the supergravity approximation is invalid near the boundary. This invalidity reflects how the BFSS matrix model is weakly coupled at short time scales. To address this issue, we apply the perturbing fields at the boundary of the supergravity region (\ref{nearboundcond}) instead of the actual boundary of spacetime at $r=\infty$, following Refs. \cite{Gubser:1998bc,Sekino:1999av,Kanitscheider:2008kd}. The lower bound of Eq. (\ref{eq:holographicrenorm}) implies this assumption is valid.

Finally, the state described by Eq. (\ref{stringframemetrci}) is technically not stable due to the emission of D0 branes by Hawking radiation. However, such rates are exponentially suppressed at large-$N$ \cite{Lin:2013jra}. Such issues do not appear for conformal holographic duals because the emitted radiation bounces back in finite time.

\subsection{The holographic map}

We now review the holographic map between type IIA supergravity and BFSS. Although the string frame metric is useful for studying the validity of the supergravity approximation, we generically perform calculations in the dual string frame for technical reasons \cite{Boonstra:1998mp,Skenderis:1998dq}. We first perform the change of variables
\begin{equation}
z=\frac{2}{5}\left (\frac{r}{(\lambda d_{0})^{1/3}}\right )^{\frac{-5}{2}}, \quad \tau=i(d_{0}\lambda)^{1/3}t
\end{equation}
so the metric becomes 
\begin{equation}\label{dualstringframemetrx}
\begin{split}
\frac{ds^{2}}{\alpha'}&=\left (\frac{5}{2}z \right )^{3/5}\left [ \left ( \frac{2}{5}\right )^{2}\left (\frac{d\tau^{2}+dz^{2}}{z^{2}}\right )+d\Omega_{8}^{2}  \right ] \ ,  \\ &e^{\phi}=\frac{(2\pi)^{2}}{d_{0}N} \left (\frac{5z}{2} \right )^{21/10} \ , \\
&A_{\tau}=-i\left ( \frac{Nd_{0}}{(2\pi)^{2}}\right )\left ( \frac{5}{2}z\right )^{-\frac{14}{5}} \ .
\end{split}
\end{equation}
Note that we have moved to Euclidean time. We can absorb the Weyl factor into the dilaton background, resulting in the dual string frame metric \cite{Boonstra:1998mp,Skenderis:1998dq}: 
\begin{equation}\label{stringframe}
\begin{split}
ds_{\textrm{dual}}^{2}&=(\frac{Nd_{0}}{(2\pi)^{2}} e^{\phi})^{\frac{-2}{7}}ds_{\textrm{string}}^{2} \\
&=\alpha' \left [ \left (\frac{2}{5}\right )^{2}\left ( \frac{d\tau^{2}+dz^{2}}{z^{2}}\right ) +d\Omega_{8}^{2}\right ]
\end{split}
\end{equation}
where $ds_{\textrm{dual}}^{2}$ is simply $\textrm{AdS}_{2}\times \textrm{S}_{8}$. Boundary correlation functions should be invariant under such field redefinitions \cite{Diwakar:2021juk}. Under this Weyl re-scaling of the metric, the relevant bosonic part of the action becomes 
\begin{equation}\label{action}
\begin{split}
\mathcal{S}_{\textrm{SUGRA}}\ni \frac{1}{(2\pi)^{7}\alpha'^{4}} \left ( \frac{Nd_{0}}{(2\pi)^{2}}  \right )^{2}\int d^{10}x \sqrt{g_{\textrm{dual}}}(\frac{Nd_{0}}{(2\pi)^{2}} &e^{\phi})^{-6/7}\bigg [R+\frac{16}{49}(\partial \phi)^{2} \\
&-\frac{1}{4}\left ( \frac{Nd_{0}}{(2\pi)^{2}}\right)^{-2}\left ( \frac{Nd_{0}}{(2\pi)^{2}} e^{\phi}\right )^{\frac{12}{7}}F_{2}^{2} \bigg ] \ .
\end{split}
\end{equation}
Note that $16\pi G_{N}=(2\pi)^{7}\alpha'^{4}$ in our conventions. We denote the boundary of the supergravity region from Eq. (\ref{nearboundcond}) as $z_{b}$, which is technically $\mathcal{O}(\lambda^{0})$. 
However, correlators are evaluated at time separations which scale as $\lambda^{1/3}$ in $z, \tau$ coordinates, see Eq. (\ref{eq:holographicrenorm}), so we can treat $z_{b}$ as parametrically small without loss of generality.

To study the bulk fields, we follow the analysis in Ref. \cite{Sekino:1999av}, which used harmonic analysis to study linearized perturbations about the metric in Eq. (\ref{stringframe}). We use $\mu,\nu,\ldots$ for 10 dimensional indices, $I,J,K,\ldots$ for metric indices of the $\textrm{S}_{8}$ sphere and $\alpha,\beta,\kappa,\ldots$ for metric indices of the AdS${}_{2}$ spacetime. For example, consider the bulk graviton\footnote{Note that this is the graviton in the dual string frame in Eq. (\ref{stringframe}), not the string frame.} with indices on the $\textrm{S}_{8}$ sphere. Working in the gauge 
\begin{equation}
\nabla_{I}\left ( h^{IJ}(x^\mu)-\frac{1}{8}g_{\textrm{S}_{8}}^{IJ} h^{K}_{K}(x^\mu) \right )=0 \ ,
\end{equation}
we decompose these components of the linear metric perturbation as 
\begin{equation}\label{tensormodes}
\begin{split}
h^{IJ}(x^\mu)-\frac{1}{8}g_{\textrm{S}_{8}}^{IJ}h^{K}_{K}(x^\mu)&=\sum_{\ell\geq 2,\vec{m}} b_{\ell,\vec{m}}(z,\tau)Y_{\ell,\vec{m}}^{IJ}(x^{K}) \ , 
\end{split}
\end{equation}
where the $Y^{IJ}_{\ell,\vec{m}}$ are the symmetric traceless tensor harmonics of $\textrm{S}_{8}$.\footnote{See Appendix A of Ref. \cite{Sekino:1999av} for details.} The $\ell$ and $\vec{m}$ indices correspond to Casimir and the Gelfand-Zetlin indices. The boundary correlators of $b_{\ell,\vec{m}}$ correspond to,
\begin{equation}\label{metricpert}
b_{\ell,\vec{m}}\leftrightarrow \textrm{STr}[(D_{t}X^{i}D_{t}X^{j}+F^{ik}F^{kj}+\frac{1}{4}\psi_{\alpha}\gamma^{i}_{\alpha\beta}[\psi_{\beta},X^{j}]+\frac{1}{4}\psi_{\alpha}\gamma^{j}_{\alpha\beta}[\psi_{\beta},X^{i}])X^{i_{1}}X^{i_{2}}\ldots X^{i_{\ell}}] 
\end{equation}
where $\textrm{STr}$ means we have symmetrized trace over all orderings of the operators \cite{Kabat:1997sa,Taylor:1998tv,Sekino:1999av,Sekino:2000mg}. Importantly, the indices on the right-hand side of Eq. (\ref{metricpert}) correspond to the S${}_{8}$ sphere instead of $\mathbb{R}^{9}$ and the trace part of the $i,j$-indices and the $i_{1}\ldots i_{\ell}$-indices should be subtracted \cite{Sekino:1999av}.\footnote{The identification in Eq. (\ref{metricpert}) can be made by considering weak field deformations on the D0-brane background before we take the decoupling limit. Such deformations modify the DBI action by a source term that can be matched to the corresponding bulk excitation in the decoupling limit \cite{Klebanov:1997kc,Klebanov:1999xv,Klebanov:2000me,Batra:2025ivy}.} See Refs. \cite{Kabat:1997sa,Taylor:1998tv,Sekino:1999av,Sekino:2000mg} for a detailed discussion of the other field content.

We are computing the leading order result in the large-$N$ approximation, which corresponds to discarding higher loop contributions in the bulk theory. Therefore, we can freely discard the heavier KK modes, which would run in loops, and focus on correlators of the lightest KK modes. The end result is a theory of scalars on a two dimensional background. We show how to compute the relevant quadratic and cubic terms of the dimensionally reduced action using xPert \cite{Brizuela:2008ra}, a Mathematica package for automating tensor algebra calculations, in Appendix \ref{quadraticcubicaction}.

We characterize fields by their behavior close to $z_{b}$:
\begin{equation}
\phi(z,\tau)\rightarrow z_{b}^{\Delta}\phi_{(0)}(\tau), \quad \textrm{as} \quad z\rightarrow z_{b} \ .
\end{equation}
where $\Delta$ corresponds to the dimension of the corresponding operator. The dimension of the $b_{\ell,\vec{m}}$ modes is
\begin{equation}\label{actualscalingdim}
\begin{split}
\Delta_{b_{\ell,\vec{m}}}&=\frac{14}{5}+\frac{2}{5}\ell \ .
\end{split}
\end{equation}
A more general discussion of the dimension of such operators is provided in Ref. \cite{Biggs:2023sqw}. To compute correlation functions, we deform the Lagrangian by source terms
\begin{equation}
S_{pert}=\int d\tau \phi_{(0)}(\tau)\mathcal{O}(\tau)
\end{equation}
and use the holographic duality 
\begin{equation}\label{holographicdiction}
\left \langle \exp \left ( \int d\tau \phi_{(0)}(\tau)\mathcal{O}(\tau) \right ) \right \rangle=e^{-\mathcal{S}_{\textrm{SUGRA}}}|_{\lim_{z\rightarrow z_{b}}\phi(z,\tau)z^{\Delta-14/5}=\phi_{(0)}(\tau)} \ ,
\end{equation}
where the above equality assumes that $z_{b}$ is much smaller than all other scales in the problem. If we assume that $\mathcal{O}(\tau)$ has dimension $\Delta$, then $\phi_{(0)}(\tau)$ has dimension $s=14/5-\Delta$. This scaling follows from how the quantum effective action of the BFSS model obeys a scaling similarity in the supergravity regime \cite{Boonstra:1998mp}:
\begin{equation}\label{similartrans}
\tau\rightarrow \gamma \tau \quad \Rightarrow \quad S\to \gamma^{-\theta}S, \quad \theta=\frac{9}{5}
\end{equation}
instead of being invariant under time re-scalings. It is easy to derive the scaling similarity in Eq. (\ref{similartrans}) from how the supergravity action behaves in Eq. (\ref{action}) under re-scaling $z\rightarrow \gamma z$ and $\tau \rightarrow \gamma \tau$. An argument for the exponent $\theta = 9/5$ from the matrix model was given in Ref. \cite{Biggs:2023sqw}. To derive the correlators, we consider functional derivatives of the right hand side of the supergravity action with sources in Eq. (\ref{holographicdiction}). 

\subsection{Two-point correlator}
\label{sec:twopointcorrelator}

In this section, we will review the computation of the bulk-to-boundary propagator and the two-point correlator \cite{Sekino:1999av,Sekino:2000mg,Kanitscheider:2008kd, Wiseman:2008qa}. We can truncate the theory to scalars on $\textrm{AdS}_{2}$ in a particular background. The most general quadratic action in the theory is 
\begin{equation}\label{twopointcorrelator}
\begin{split}
\mathcal{S}_{\textrm{SUGRA}}\ni C_{\phi_{i}^{2}}\int d\tau dz z^{-9/5}[\partial_{z}\phi(z,\tau)\partial_{z}\phi(z,\tau)+\partial_{\tau}\phi(z,\tau)\partial_{\tau}\phi(z,\tau)+\frac{m^{2}}{z^{2}}\phi(z,\tau)^{2}]
\end{split}
\end{equation}
where $C$ is some constant. For $\phi=b_{\ell,\vec{m}}$, we find that 
\begin{equation}
\begin{split}
C_{b_{\ell,\vec{m}}^{2}}&=\frac{1}{(2\pi)^{7}}\left (\frac{N d_{0}}{(2\pi)^{2}} \right )^{2}\left ( \frac{2}{5}\right)^{19/5}\left (\frac{25}{8} \right ) \ , \\
&\textrm{assuming} \quad 1=\int d\Omega_{8} Y_{\ell,\vec{m}}^{IJ}Y_{\ell,\vec{m},IJ} \ ,
\end{split}
\end{equation}
and 
\begin{equation}
m^{2}=\frac{4\ell(\ell+7)}{25} \ .
\end{equation}
We solve the equations of motion for the bulk-to-boundary propagator with the ansatz
\begin{equation}\label{bulktoboundaryunnorm}
K(z,\tau;\tau')\propto \frac{z^{\Delta}}{(z^{2}+(\tau-\tau')^{2})^{\Delta-9/10}} \ ,
\end{equation}
finding 
\begin{equation}
m^{2}=\Delta (\Delta-\frac{14}{5}) \ .
\end{equation}
We discarded the other solution to the equations of motion by imposing that the solution is regular in the interior. The BFSS bulk-to-boundary propagator in Eq.~(\ref{bulktoboundaryunnorm}) differs from the standard AdS bulk-to-boundary propagator due to the presence of the dilaton background and the Weyl factor in the metric. To normalize the bulk-to-boundary propagator, we require
\begin{equation}\label{boundarycond}
\lim_{z_{b}\rightarrow 0} \lim_{z\rightarrow z_{b}}z_{b}^{\Delta-14/5}K(z,\tau;\tau')=\delta(\tau-\tau')
\end{equation}
which leads to 
\begin{equation}\label{bfssbulkbound}
K(z,\tau;\tau')=\frac{\Gamma[\Delta-9/10]}{\sqrt{\pi}\Gamma[\Delta-14/10]}\frac{z^{\Delta}}{(z^{2}+(\tau-\tau')^{2})^{\Delta-9/10}} \ .
\end{equation}
Eq. (\ref{boundarycond}) comes from how we impose the boundary conditions at the boundary of the supergravity region at $z=z_{b}$, not $z=0$. As previously discussed, we should technically not take $z_{b}$ to $0$ in Eq. (\ref{boundarycond}) because $z_{b}$ is $\mathcal{O}(1)$ in string units. We again emphasize that all other scales in the problem are parametrically larger than $z_{b}$, so we can take $z_{b}\rightarrow 0$ without loss of generality.

We now compute the boundary-to-boundary two-point function. We will work entirely in position space. We compute the action as a functional of two source fields
\begin{equation}
\begin{split}
S[\phi_{(0)}(\tau)]&\supseteq C_{\phi_{i}^{2}}\int d\tau dz z^{-9/5}[\partial_{z}\phi(z,\tau)\partial_{z}\phi (z,\tau) \\
&+\partial_{\tau}\phi(z,\tau)\partial_{\tau}\phi(z,\tau)+\frac{m^{2}}{z^{2}}\phi\phi] \\
&=C_{\phi_{i}^{2}}\int d\tau [z^{-9/5}\phi\partial_{z}\phi]_{z=z_{b}}^{\infty}
\end{split}
\end{equation}
where
\begin{equation}
\phi(z,\tau)=\int d\tau' K(z,\tau;\tau')\phi_{(0)}(\tau') 
\end{equation}
and we need to take the leading contribution in the $z_{b}\rightarrow 0$ limit. In principle, the action contains higher order terms, but these do not contribute to the two-point function. The two-point function is given by 
\begin{equation}\label{almostlaurent}
\begin{split}
\langle \mathcal{O}(\tau_{1})\mathcal{O}(\tau_{2})\rangle &=-\frac{\delta^{2}S}{\delta \phi_{(0)}(\tau_{1})\delta \phi_{(0)}(\tau_{2})} \\
&=\lim_{z_{b}\rightarrow 0}C_{\phi_{i}^{2}} \frac{2\Gamma \left(\Delta -\frac{9}{10}\right)^2 z_b^{2\Delta-\frac{14}{5}}}{5 \pi  \Gamma \left(\Delta -\frac{7}{5}\right)^2} \int d\tau \left(z_b^2+\left(\tau -\tau _1\right){}^2\right){}^{\frac{9}{10}-\Delta } \\ 
&\times \left(z_b^2+\left(\tau -\tau _2\right){}^2\right){}^{-\Delta -\frac{1}{10}}
[(9-5 \Delta ) z_b^2+5 \Delta  \left(\tau -\tau _2\right){}^2] \ .
\end{split}
\end{equation}
We now use the identity %
\begin{equation}\label{eq:crucialiden2p}
\begin{split}
&\lim_{z_{b}\rightarrow 0} \int d\tau \left(z_b^2+\left(\tau-\tau_1\right){}^2\right){}^{\frac{9}{10}-\Delta } \left(z_b^2+\left(\tau-\tau_2\right){}^2\right){}^{-\Delta -\frac{1}{10}}[(9-5 \Delta ) z_b^2+5 \Delta  \left(\tau-\tau_2\right){}^2]\\
&=\frac{14 \sqrt{\pi } \Gamma \left(\Delta -\frac{7}{5}\right) z_b^{\frac{14}{5}-2 \Delta }}{\Gamma \left(\Delta -\frac{9}{10}\right)\left|\tau _1-\tau _2\right|{}^{2 \Delta-\frac{9}{5} } }
\end{split}
\end{equation}
which is derived in Appendix \ref{derivationequ} using the techniques in Appendix \ref{intromethodregions}.\footnote{Note that Eq. (\ref{almostlaurent}) is not an Euler integral because the polynomials in the integrand are not Laurent polynomials.} The final result is 
\begin{equation}\label{norm2point}
\langle \mathcal{O}(\tau_{1})\mathcal{O}(\tau_{2})\rangle=C_{\phi_{i}^{2}}\frac{28  \Gamma \left(\Delta -\frac{9}{10}\right) }{5 \sqrt{\pi } \Gamma \left(\Delta -\frac{7}{5}\right)}\frac{1}{|\tau _1-\tau _2|^{2 \Delta-9/5 }}
\end{equation}
which matches the expressions computed in Refs. \cite{Sekino:1999av,Sekino:2000mg} up to a normalization factor. We normalize the boundary operators so that they obey Eq. (\ref{norm2point}). That is, given the right-hand side of Eq. (\ref{metricpert}) in the boundary theory times a normalization constant, one chooses the constant such that Eq. (\ref{norm2point}) holds. It would be desirable to compute this normalization constant in the boundary theory, such as by using supersymmetry \cite{Asano:2012zt,Asano:2014vba,Asano:2017xiy}.\footnote{We thank Juan Maldacena for discussion on this point.} 
We note that a direct computation of the two-point function in the IIA gravity regime of BFSS using supersymmetry is subtle. The IIA supergravity description applies to near-extremal D0-brane black holes at finite temperature, but the thermal circle with anti-periodic fermion boundary conditions breaks all sixteen supercharges, precluding a straightforward application of localization. Nevertheless, in the near-extremal limit $T << \lambda^{1/3}$ this breaking is a soft effect controlled by the small parameter $T^3/\lambda$, and it is natural to ask whether localization, combined with a systematic expansion in $T^3/\lambda$, could yield exact results for the leading corrections to protected correlators. We leave a detailed investigation of such near-BPS localization schemes to future work.

\section{Three-point correlator}\label{threepointcorrelator}

We now consider the computation of a three-point function. The relevant cubic interaction term in the action takes the form
\begin{equation}
\begin{split}
S\supseteq &\ C^{(1)}_{\phi_{1}\phi_{2}\phi_{3}}\int d\tau dz z^{-19/5}\phi_{1}(z,\tau)\phi_{2}(z,\tau)\phi_{3}(z,\tau)\ . 
\end{split}
\end{equation}
The procedure for computing $C^{(i)}_{\phi_{1}\phi_{2}\phi_{3}}$ for the $b_{\ell,\vec{m}}(z,\tau)$ modes in terms of integrals over spherical harmonics is given in Appendix \ref{quadraticcubicaction}. The relevant Witten diagram gives the contribution
\begin{equation}\label{threepoint}
\begin{split}
W_{3}(\tau_{i},\Delta_{i})&=S_{\phi_{1}\phi_{2}\phi_{3}}C^{(1)}_{\phi_{1}\phi_{2}\phi_{3}}\times \begin{tikzpicture}[baseline={([yshift=-.5ex]current bounding box.center)},every node/.style={font=\scriptsize}]\pgfmathsetmacro{\r}{0.8}
\draw [] (0,0) circle (\r cm);
\tikzset{decoration={snake,amplitude=.4mm,segment length=1.5mm,post length=0mm,pre length=0mm}}
\filldraw (0:\r) circle (1pt) node[right=0pt]{$1$};
\filldraw (120:\r) circle (1pt) node[above=0pt]{$2$};
\filldraw (240:\r) circle (1pt) node[below=0pt]{$3$};
\draw [thick] (0:\r) -- (0:0);
\draw [thick] (120:\r) -- (120:0);
\draw [thick] (240:\r) -- (240:0);
\end{tikzpicture} \\
&=S_{\phi_{1}\phi_{2}\phi_{3}}C^{(1)}_{\phi_{1}\phi_{2}\phi_{3}} \int d\tau dz z^{-19/5} \prod_{i=1}^{3}\frac{\Gamma(\Delta_{i}-9/10)}{\sqrt{\pi}\Gamma(\Delta_{i}-14/10)}\frac{z^{\Delta_{i}}}{[z^{2}+(\tau-\tau_{i})^{2}]^{\Delta_{i}-9/10}}
\end{split}
\end{equation}
where $S_{\phi_{1}\phi_{2}\phi_{3}}$ is a symmetry factor.  

Let us first consider when the leading Witten diagram in Eq. (\ref{threepoint}) is a valid approximation of the three-point correlator. In general, Monte-Carlo cannot study arbitrarily low-temperature systems, which is in conflict with the low-temperature assumption in Eq. (\ref{assumption3}). Nonetheless, as emphasized in the introduction, the low-temperature assumption is still valid for computing correlators at finite temperatures as long as the operators are sufficiently localized. At finite temperature $T$, the horizon is located at 
\begin{equation}
z_{h}\sim \lambda^{1/3}T^{-1}
\end{equation}
in $z,\tau$-coordinates. From a saddle point analysis of the $z$ integral, the leading Witten diagram in Eq. (\ref{threepoint}) is dominated by interactions at $z \lesssim \textrm{max}(|\tau_{12}|,|\tau_{13}|,|\tau_{23}|)$, where $\tau_{ij}=|\tau_{i}-\tau_{j}|$. Therefore, the validity of the low-temperature assumption in Eq. (\ref{assumption3}) requires that 
\begin{equation}\label{validitiapproxatio}
\forall \ i,j:\quad \lambda^{1/3}T^{-1}>>|\tau_{ij}| \ .
\end{equation}
In addition to Eq. (\ref{validitiapproxatio}), there is a lower bound on $|\tau_{ij}|$ due to the large curvature of the near-boundary region, as discussed in Section \ref{gravitydual}. It is in the intermediate regime given in the introduction, see Eq. (\ref{eq:holographicrenorm}), that the leading Witten diagram is a good approximation of the three-point correlator. 

\subsection{The three-point Witten diagram as an Euler integral}\label{eulerintegral}

We now convert this Witten diagram to an Euler integral. We first apply the Schwinger trick
\begin{equation}\label{swchingerrepre}
\frac{1}{[z^{2}+(\tau-\tau_{i})^{2}]^{n_{i}}}=\frac{1}{\Gamma(n_{i})}\int d\alpha_{i}\alpha_{i}^{n_{i}-1}e^{-\alpha_{i}[z^{2}+(\tau-\tau_{i})^{2}]}
\end{equation}
and then evaluate the $\tau$ and $z$ integrals using Gaussian integration. The end result is the ``Schwinger" representation of the integral
\begin{equation}
\begin{split}
&W_{3}(\tau_{i},\Delta_{i})=S_{\phi_{1}\phi_{2}\phi_{3}}C^{(1)}_{\phi_{1}\phi_{2}\phi_{3}}\frac{\Gamma \left(\frac{1}{2} \left(\Delta _{123}-\frac{14}{5}\right)\right)}{2 \pi  \Gamma \left(\Delta _1-\frac{7}{5}\right) \Gamma \left(\Delta _2-\frac{7}{5}\right) \Gamma \left(\Delta _3-\frac{7}{5}\right)} \int [\prod_{i=1}d\alpha_{i}\alpha _i^{\Delta _i-\frac{19}{10}}]\\
&\quad \times \left(\alpha _1+\alpha _2+\alpha _3\right){}^{\frac{1}{2}[\frac{9}{5}-\sum_{i}\Delta_{i}]} \exp \left(-\frac{\alpha _2 \alpha _3 \tau_{23}^{2}+\alpha _1 \alpha _2 \tau_{12}^{2}+\alpha _1\alpha _3 \tau_{13}^{2}}{\alpha _1+\alpha _2+\alpha _3}\right) \ ,
\end{split}
\end{equation}
where
\begin{equation}
\Delta_{123}=\Delta_{1}+\Delta_{2}+\Delta_{3} \ .
\end{equation}
We then convert to the ``Feynman" representation of the integral. We introduce the delta function 
\begin{equation}
1=\int d\lambda \delta(\lambda -\sum \alpha_{i})
\end{equation}
re-scale all the $\alpha$-parameters, $\alpha_{i}\rightarrow \lambda \alpha_{i}$, and then integrate over $\lambda$. Following Ref. \cite{Hillman:2023ezp}, we finally convert to an Euler integral by making the substitution
\begin{equation}
u_{1}=\frac{\alpha_{1}}{\alpha_{3}}, \quad u_{2}=\frac{\alpha_{2}}{\alpha_{3}}
\end{equation}
which leads to the final form of the integral, 
\begin{equation}\label{3pointfinal}
\begin{split}
W_{3}(\tau_{i},\Delta_{i})&=\frac{S_{\phi_{1}\phi_{2}\phi_{3}}C^{(1)}_{\phi_{1}\phi_{2}\phi_{3}}\ \Gamma \left(\frac{1}{2} \left(\Delta _{123}-\frac{14}{5}\right)\right) \Gamma \left(\frac{1}{2}\Delta _{123}-\frac{9}{5}\right)}{(2\pi)\prod_{i}\Gamma[\Delta_{i}-7/5]} \\
&\times \tau _{12}^{-2t_{2}}\int du_{1} du_{2} u_{1}^{\Delta_{1}-19/10}u_{2}^{\Delta_{2}-19/10}p_{1}^{-t_{1}}p_{2}^{-t_{2}}
\end{split}
\end{equation}
where
\begin{equation}\label{eulerrep1}
\begin{split}
p_{1}&=1+u_{1}+u_{2}, \quad t_{1}=9/10 \ , \\
p_{2}&= u_1 u_2+\frac{\tau _{13}^2}{\tau _{12}^2} u_1+\frac{\tau _{23}^2}{\tau _{12}^2}  u_2, \quad t_{2}=\frac{1}{2}\Delta _{123}-\frac{9}{5} \ . 
\end{split}
\end{equation}
Unfortunately, we could not find a nice closed-form expression for this integral. However, the Euler integral representation of the three-point correlator allows us to apply powerful mathematical techniques. Alternatively, one could simply evaluate the integral numerically. In this letter, we focus on analytically evaluating the correlator in particular limits using the method of regions.

\begin{figure}[t]
\centering
\includegraphics[width=8cm]{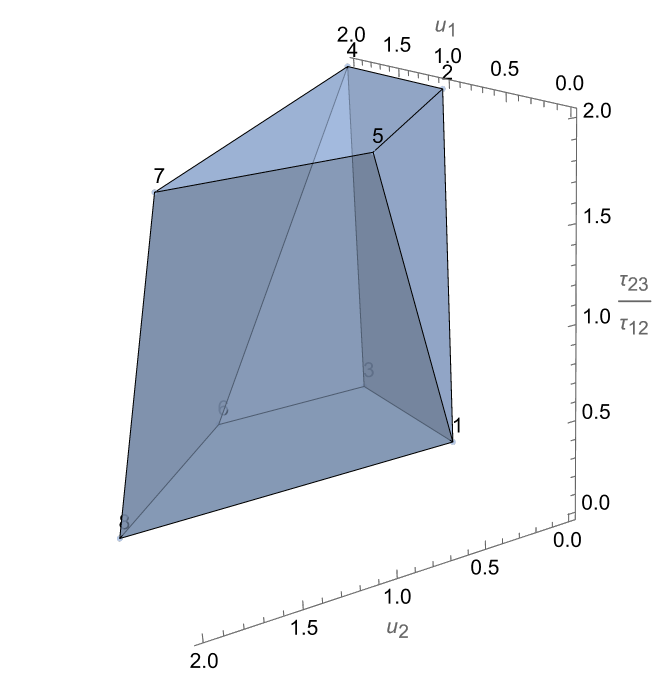}
\caption{The Newton polytope of the product of polynomials $p_{1}$ and $p_{2}$ in Eq. (\ref{eulerrep1}).}\label{fignewtonreigon1}
\end{figure}

\subsection{The squeezed limit}

We now assume that $\tau_{1}>\tau_{2}>\tau_{3}$ and substitute $\tau_{13}=\tau_{12}+\tau_{23}$ in order to study the squeezed limit that 
\begin{equation}
\begin{split}
&\frac{\tau_{23}}{\tau_{12}} << 1 
\end{split}
\end{equation}
where we take $\tau_{12}$ to be fixed. We apply the method of regions, which is reviewed in Appendix \ref{intromethodregions}, to compute the integral in this limit. The Newton polytope of 
\begin{equation}
p_{1}p_{2}=(1+u_{1}+u_{2})\left (u_1 u_2+\left (1+\frac{\tau _{23}}{\tau _{12}} \right )^2 u_1+\frac{\tau _{23}^2}{\tau _{12}^2}  u_2 \right )
\end{equation}
with respect to the variables $(u_{1},u_{2},\tau_{23}/\tau_{12})$ has the vertices
\begin{equation}
\begin{split}
\textrm{Vertices}:\quad (1,0,0),\ (1,0,2), \ (2,0,0),\ (2,0,2), \ (0,1,2), \ (2,1,0), \ (0,2,2), \ (1,2,0)
\end{split}
\end{equation}
and is given in Fig. \ref{fignewtonreigon1}. The facet vectors are given by 
\begin{equation}
\begin{split}
\textrm{Facets}:&\quad (2, 2, 0),\ (-2, 0, 0),\ (2, 0, 1),\ (0, 0, -1),\ (0, 2, 0),\ (0, 0, 2), \
(-2, -2, -1) \ .
\end{split}
\end{equation}
There are two downward pointing facets along the axis corresponding to $\tau_{23}/\tau_{12}$:
\begin{equation}
\begin{split}
\vec{\gamma}_{1}&=(0,0,1) \\
\vec{\gamma}_{2}&=(2,0,1) 
\end{split}
\end{equation}
which correspond to the regions
\begin{equation}\label{regioncon}
\begin{split}
\textrm{Region 1}:& \quad u_{1}\rightarrow \left ( \frac{\tau_{23}}{\tau_{12}} \right )^{0}u_{1}, \quad u_{2}\rightarrow \left ( \frac{\tau_{23}}{\tau_{12}} \right )^{0}u_{2} \ ,\\
\textrm{Region 2}:& \quad u_{1}\rightarrow \left ( \frac{\tau_{23}}{\tau_{12}} \right )^{2}u_{1}, \quad u_{2}\rightarrow \left ( \frac{\tau_{23}}{\tau_{12}} \right )^{0}u_{2} \ .
\end{split}
\end{equation}
To derive the leading contribution from a given region, we re-scale the integrand according to Eq. (\ref{regioncon}) and then series expand the integrand in $\tau_{23}/\tau_{12}$. The leading contributions from Region 1 is 
\begin{equation}\label{3pointfinalreg1}
\begin{split}
W_{3}(\tau_{i},\Delta_{i})&\ni \frac{S_{\phi_{1}\phi_{2}\phi_{3}}C^{(1)}_{\phi_{1}\phi_{2}\phi_{3}}\ \Gamma \left(\frac{1}{2} \left(\Delta _{123}-\frac{14}{5}\right)\right) \Gamma \left(\frac{1}{2}\Delta _{123}-\frac{9}{5}\right)}{(2\pi)\prod_{i}\Gamma[\Delta_{i}-7/5]} \\
&\times \tau _{12}^{-2t_{2}}\int du_{1} du_{2}u_{1}^{\Delta_{1}-19/10}u_{2}^{\Delta_{2}-19/10}p_{1}'^{-t_{1}}p_{2}'^{-t_{2}}
\end{split}
\end{equation}
where
\begin{equation}
\begin{split}
p_{1}'&= 1+u_{1}+u_{2}   \ , \\
p_{2}'&= u_{1}(1+u_{2})   \ .
\end{split}
\end{equation}
The leading contributions from Region 2 is
\begin{equation}\label{3pointfinalreg2}
\begin{split}
W_{3}(\tau_{i},\Delta_{i})&\ni \frac{S_{\phi_{1}\phi_{2}\phi_{3}}C^{(1)}_{\phi_{1}\phi_{2}\phi_{3}}\ \Gamma \left(\frac{1}{2} \left(\Delta _{123}-\frac{14}{5}\right)\right) \Gamma \left(\frac{1}{2}\Delta _{123}-\frac{9}{5}\right)}{(2\pi)\prod_{i}\Gamma[\Delta_{i}-7/5]} \\
&\times \tau _{12}^{-2t_{2}}\left ( \frac{\tau _{23}}{\tau _{12}} \right )^{ \Delta _1-\Delta _2- \Delta _3+\frac{9}{5}}\int du_{1} du_{2} u_{1}^{\Delta_{1}-19/10}u_{2}^{\Delta_{2}-19/10}p_{1}''^{-t_{1}}p_{2}''^{-t_{2}}
\end{split}
\end{equation}
where
\begin{equation}
\begin{split}
p_{1}''&= 1+u_{2}  \ , \\
p_{2}''&= u_{1}+u_{2}+u_{1}u_{2}   \ .
\end{split}
\end{equation}
It is now clear that if 
\begin{equation}
\Delta _1+\frac{9}{5}<\Delta _2+\Delta _3
\end{equation}
then Eq. (\ref{3pointfinalreg2}) will dominate at small $\tau_{23}/\tau_{12}$, implying the scaling behavior (\ref{scalebehavior}) quoted in the introduction.

We can also compute the exact numeric coefficient. Unlike the original integral in Eq. (\ref{3pointfinal}), the integrals in Eqs. (\ref{3pointfinalreg1}) and (\ref{3pointfinalreg2}) evaluate to numeric coefficients for fixed $\Delta_{i}$. In fact, both integrals can be evaluated in terms of Gamma functions by evaluating the $u_{1}$ integral first and then the $u_{2}$ integral:
\begin{equation}
\begin{split}
\int du_{1}du_{2} u_{1}^{\Delta_{1}-19/10}u_{2}^{\Delta_{2}-19/10}p_{1}'^{-t_{1}}p_{2}'^{-t_{2}}&=\frac{\Gamma \left(\Delta _2-\frac{9}{10}\right) \Gamma \left(\Delta _3-\frac{9}{10}\right) \Gamma \left(\frac{9}{10}-\frac{\Delta _{\text{23,1}}}{2}\right) \Gamma \left(\frac{\Delta _{\text{23,1}}}{2}\right)}{\Gamma \left(\frac{9}{10}\right) \Gamma \left(\Delta _2+\Delta _3-\frac{9}{5}\right)}\\
\int du_{1}du_{2} u_{1}^{\Delta_{1}-19/10}u_{2}^{\Delta_{2}-19/10}p_{1}''^{-t_{1}}p_{2}''^{-t_{2}}&=\frac{\Gamma \left(\Delta _1-\frac{9}{10}\right) \Gamma \left(\frac{\Delta _{\text{12,3}}}{2}\right) \Gamma \left(\frac{\Delta _{\text{13,2}}}{2}\right) \Gamma \left(\frac{1}{2}\Delta _{\text{23,1}}-\frac{9}{10} \right)}{\Gamma \left(\Delta _1\right) \Gamma \left(\frac{1}{2} \Delta _{123}-\frac{9}{5}\right)}\\
\end{split}
\end{equation}
where
\begin{equation}
\begin{split}
\Delta_{ij,k}&=\Delta_{i}+\Delta_{j}-\Delta_{k} \ .
\end{split}
\end{equation}

\subsection{Comparison with uplift to AdS${}_{2+9/5}$}\label{dilatonuplift}

Refs. \cite{Kanitscheider:2009as,Biggs:2023sqw} have considered an alternative approach to dealing with the running dilaton. The idea is to treat the $d=10$ action with a non-trivial dilaton profile as coming from a $d=10+\bar{\theta}$ action where the original dilaton is the volume of the $\bar\theta$-dimensional auxiliary spacetime. For the background metric in Eq. (\ref{dualstringframemetrx}), the uplifted metric after performing the dimensional reduction on S${}_{8}$ is 
\begin{equation}
ds^{2}\propto \frac{dz^{2}+d\tau^{2}+dx_{\bar{\theta}}^{2}}{z^{2}}
\end{equation}
where $\bar{\theta}=9/5$. This uplift should not be accurate beyond tree level because of contamination from unphysical KK modes of the uplifted dimension. However, it is sufficient for tree-level calculations and classical gravity. For example, Ref. \cite{Biggs:2023sqw} used the uplifted metric to compute the quasi-normal modes. Upon implementing the uplift, correlators are integrated along the $\bar{\theta}$ dimension. Suppose we are computing boundary correlators of the operator $\mathcal{O}(x)$ with dimension $\Delta$. Using the uplift, the correlator can be re-written in terms of correlators in a conformal $d=1+\bar\theta$ dimensional theory with conformal dimension $\Delta$:
\begin{equation}
\mathcal{O}=\int d^{\bar{\theta}}x_{\bar{\theta}} \mathcal{O}_{\textrm{uplift}}(x+x_{\bar{\theta}})
\end{equation}
where we also need to mod the entire correlator by a translation symmetry in the $x_{\bar\theta}$ direction. 

For example, consider the two-point correlator, which was computed using this uplift procedure in Ref. \cite{Biggs:2023sqw}. Using the above prescription, the uplifted correlator can be written as 
\begin{equation}
\langle \mathcal{O}(\tau_{1})\mathcal{O}(\tau_{2})\rangle\propto \int \frac{d^{\bar{\theta}}x_{1,\bar{\theta}}d^{\bar{\theta}}x_{2,\bar{\theta}}}{\textrm{Vol}(\mathbb{R}^{\bar\theta})} \langle \mathcal{O}_{\textrm{uplift}}(\tau_{1}+x_{1,\bar{\theta}})\mathcal{O}_{\textrm{uplift}}(\tau_{2}+x_{2,\bar{\theta}})\rangle \ .
\end{equation}
where the $\textrm{Vol}(\mathbb{R}^{\bar\theta})$ corresponds to the translation symmetry in the auxiliary dimension. We take the operator to have conformal dimension $\Delta$. We use the translation symmetry to fix $x_{2,\bar{\theta}}=0$, finding
\begin{equation}\label{twopointresult}
\begin{split}
\langle \mathcal{O}(\tau_{1})\mathcal{O}(\tau_{2})\rangle&\propto  \int d^{\bar{\theta}}x_{1,\bar{\theta} }\frac{1}{(\tau_{12}^{2}+x_{1,\bar{\theta}}^{2})^{\Delta}} \\
&\propto \frac{1}{\tau_{12}^{2\Delta-\bar{\theta}}} \ .
\end{split}
\end{equation}
Eq. (\ref{twopointresult}) matches the computation in Section \ref{sec:twopointcorrelator}.

We can now consider the three-point correlator. Using the uplift prescription and how the kinematic dependence of three-point boundary correlators is fixed by the bulk isometry of $\textrm{AdS}_{2+\bar\theta}$, we find 
\begin{equation}
\langle \mathcal{O}_{\Delta_{1}}(\tau_{1})\mathcal{O}_{\Delta_{2}}(\tau_{2})\mathcal{O}_{\Delta_{3}}(\tau_{3})\rangle\propto \int  \frac{d^{\bar{\theta}}x_{1,\bar{\theta}}d^{\bar{\theta}}x_{2,\bar{\theta}}}{(\tau_{12}^{2}+(x_{1,\bar{\theta}}-x_{2,\bar{\theta}})^{2})^{\frac{\Delta_{12,3}}{2}}(\tau_{13}^{2}+x_{1,\bar{\theta}}^{2})^{\frac{\Delta_{13,2}}{2}}(\tau_{23}^{2}+x_{2,\bar{\theta}}^{2})^{\frac{\Delta_{23,1}}{2}}}
\end{equation}
We follow Section \ref{eulerintegral} by again applying Schwinger's trick, Eq. (\ref{swchingerrepre}), to each of the factors in the denominator and then integrating over $x_{i,\bar\theta}$:
\begin{equation}
\begin{split}
&\langle \mathcal{O}_{\Delta_{1}}(\tau_{1})\mathcal{O}_{\Delta_{2}}(\tau_{2})\mathcal{O}_{\Delta_{3}}(\tau_{3})\rangle \propto \int \frac{da_{1,2}}{a_{1,2}}\frac{da_{1,3}}{a_{1,3}}\frac{da_{2,3}}{a_{2,3}}a_{1,2}^{\frac{1}{2} \Delta _{12,3}} a_{1,3}^{\frac{1}{2} \Delta _{13,2}} a_{2,3}^{\frac{1}{2} \Delta _{23,1}} 
 \\
 &\quad \times \left(a_{1,3} a_{2,3}+a_{1,2} \left(a_{1,3}+a_{2,3}\right)\right){}^{-\frac{\bar\theta }{2}} \exp \left(-\tau _{12}^2 a_{1,2}-\tau _{13}^2 a_{1,3}-\tau _{23}^2 a_{2,3}\right) \ .
\end{split}
\end{equation}
In order to match the result in Section \ref{eulerintegral}, we first make the change of variables 
\begin{equation}
a_{i,j}\rightarrow \alpha_{i}\alpha_{j} \ ,
\end{equation}
insert the delta function 
\begin{equation}
1=\int d\lambda \delta(\lambda -\alpha_{1}-\alpha_{2}-\alpha_{3}) 
\end{equation}
perform the re-scaling $\alpha_{i}\rightarrow \lambda \alpha_{i}$, and finally integrate over $\lambda$. The result is 
\begin{equation}
\begin{split}
&\langle \mathcal{O}_{\Delta_{1}}(\tau_{1})\mathcal{O}_{\Delta_{2}}(\tau_{2})\mathcal{O}_{\Delta_{3}}(\tau_{3})\rangle \propto \int \left [\prod_{i=1}^{3}\frac{d\alpha_{i}}{\alpha_{i}} \alpha_{i}^{\Delta_{i}-\frac{\bar\theta}{2}}\right ](\sum_{i<j}\alpha_{i}\alpha_{j}\tau_{ij}^{2})^{\frac{1}{2}\left (2\bar\theta-\sum_{i}\Delta_{i} \right )} \delta(1-\sum_{i}\alpha_{i})
\end{split}
\end{equation}
which is proportional to the previous result in Eq. (\ref{3pointfinal}) under the change of variables $u_{1}=\alpha_{1}/\alpha_{3}$ and $u_{2}=\alpha_{2}/\alpha_{3}$ and identifying $\bar\theta=9/5$. Therefore, we have verified that the uplift procedure reproduces the desired three-point result up to a numeric coefficient.

\section{Higher point correlators}

We can now turn to computing higher point correlators. The bulk-to-bulk propagator is defined by the differential equation 
\begin{equation}
\begin{split}
&\frac{1}{e^{-6\phi/7}\sqrt{g}}\frac{\partial}{\partial x^{\alpha}} \left ( e^{-6\phi/7}\sqrt{g} g^{\alpha\beta}\frac{\partial}{\partial x^{\beta}} G(z,\tau;z',\tau')\right )-\left ( \frac{5}{2}\right )^{2}m^{2}G(z,\tau;z',\tau') \\
&=\frac{\delta(z-z')\delta(\tau-\tau')}{e^{-6\phi/7}\sqrt{g}} \ .
\end{split}
\end{equation}
The numeric factor in front of the $m^{2}$ term is to match Eq. (\ref{twopointcorrelator}).  Inputting the dilaton background, we find the second-order differential equation 
\begin{equation}\label{greensfunction}
[z^{2}\partial_{z}^{2}+z^2\partial_{\tau}^{2}-\frac{9}{5}z\partial_{z}-m^{2}]G(z,\tau;z',\tau')= C' z^{19/5}\delta(\tau-\tau')\delta(z-z')
\end{equation}
where $C'=(5/2)^{9/5}((2\pi)^{2}/(Nd_{0}))^{6/7}$ is a normalization factor. To derive the Green's function, we use a particular spectral representation of the delta-function and Green's function \cite{Penedones:2007ns,Cornalba:2007fs,Cornalba:2008qf}
\begin{equation}\label{spectralbuble}
\begin{split}
z^{19/5}\delta(\tau-\tau')\delta(z-z')&=(zz')^{9/10}\int \frac{dc}{2\pi i}\frac{\Gamma(\frac{1}{2}+c)\Gamma(\frac{1}{2}-c)}{2\pi\Gamma(c)\Gamma(-c)} \\
&\times \int d\tau'' \left ( \frac{z}{z^{2}+(\tau-\tau'')^{2}}\right )^{\frac{1}{2}+c}\left ( \frac{z'}{z'^{2}+(\tau''-\tau')^{2}}\right )^{\frac{1}{2}-c} \ .
\end{split}
\end{equation}
Plugging the spectral form of the delta function (\ref{spectralbuble}) into Eq. (\ref{greensfunction}), we find the spectral representation of bulk-to-bulk two-point propagator in BFSS
\begin{equation}
\begin{split}
G(z,\tau;z',\tau')&= \frac{C'}{[z^{2}\partial_{z}^{2}+z^2\partial_{\tau}^{2}-\frac{9}{5}z\partial_{z}-m^{2}]} z^{19/5}\delta(\tau-\tau')\delta(z-z') \\
&=C'(zz')^{9/10}\int \frac{dc}{2\pi i}\frac{\Gamma(\frac{1}{2}+c)\Gamma(\frac{1}{2}-c)}{2\pi\Gamma(c)\Gamma(-c)} \frac{1}{c^{2}-(7/5)^{2}-m^{2}} \\
&\times \int d\tau'' \left ( \frac{z}{z^{2}+(\tau''-\tau)^{2}}\right )^{\frac{1}{2}+c}\left ( \frac{z'}{z'^{2}+(\tau''-\tau')^{2}}\right )^{\frac{1}{2}-c} \ .
\end{split}
\end{equation}
Note that the jump from the first to the second line follows from how the integrand is an eigenfunction of the Laplacian 
\begin{equation}
[z^{2}\partial_{z}^{2}+z^2\partial_{\tau}^{2}-\frac{9}{5}z\partial_{z}-m^{2}]^{A} \frac{z^{\frac{7}{5}+c}}{[z^{2}+(\tau''-\tau)^{2}]^{1/2+c}} =[c^{2}-(7/5)^{2}-m^{2}]^{A}\frac{z^{\frac{7}{5}+c}}{[z^{2}+(\tau''-\tau)^{2}]^{1/2+c}} \ .
\end{equation}
for constant $A$. Identifying $m^{2}=\Delta(\Delta-14/5)$ and using the spectral represention of the AdS${}_{2}$ two-point function, 
\begin{equation}\label{BFssbulkbulkprop}
\begin{split}
G^{\textrm{AdS}_{2}}_{\Delta}(z,\tau;z',\tau')&=\int \frac{dc}{2\pi i}\frac{\Gamma(\frac{1}{2}+c)\Gamma(\frac{1}{2}-c)}{2\pi\Gamma(c)\Gamma(-c)}\times \frac{1}{c^{2}-(\Delta-1/2)^{2}} \\
& \times \int d\tau'' \left ( \frac{z}{z^{2}+(\tau-\tau'')^{2}}\right )^{\frac{1}{2}+c}\left ( \frac{z'}{z'^{2}+(\tau''-\tau')^{2}}\right )^{\frac{1}{2}-c} \ , 
\end{split}
\end{equation}
it is clear that the BFSS propagators can be written in terms of AdS${}_{2}$ propagators:
\begin{equation}\label{dimensionrel}
\begin{split}
K^{\textrm{BFSS}}_{\Delta}(z,\tau;\tau')&\propto z^{9/10}K^{\textrm{AdS}_{2}}_{\Delta-9/10}(z,\tau;\tau') \\
G^{\textrm{BFSS}}_{\Delta}(z,\tau;z',\tau')&\propto (zz')^{9/10}G^{\textrm{AdS}_{2}}_{\Delta-9/10}(z,\tau;z',\tau') 
\end{split}
\end{equation}
and that the BFSS bulk Witten diagrams are simply AdS${}_{2}$ diagrams with additional $z$ factors at each vertex. For example, the four-point tree level Witten diagram corresponds to
\begin{equation}
\begin{split}
\begin{tikzpicture}[baseline={([yshift=-.5ex]current bounding box.center)},every node/.style={font=\scriptsize}]\pgfmathsetmacro{\r}{0.8}
\draw [] (0,0) circle (\r cm);
\tikzset{decoration={snake,amplitude=.4mm,segment length=1.5mm,post length=0mm,pre length=0mm}}
\filldraw (-45:\r) circle (1pt) node[right=0pt]{$\Delta_{1}$};
\filldraw (45:\r) circle (1pt) node[right=0pt]{$\Delta_{2}$};
\filldraw (135:\r) circle (1pt) node[left=0pt]{$\Delta_{3}$};
\filldraw (225:\r) circle (1pt) node[left=0pt]{$\Delta_{4}$};
\filldraw (-90:\r/4)  node[]{$\Delta_{I}$};
\filldraw (0:\r/2) circle (0pt) node{};
\filldraw (180:\r/2) circle (0pt) node{};
\draw [thick] (-45:\r) -- (0:\r/2);
\draw [thick] (45:\r) -- (0:\r/2);
\draw [thick] (135:\r) -- (180:\r/2);
\draw [thick] (225:\r) -- (180:\r/2);
\draw [thick] (180:\r/2) -- (0:\r/2);
\end{tikzpicture}&\propto \int \frac{dz_{a}d\tau_{a}dz_{b}d\tau_{b}}{z_{a}^{11/10}z_{b}^{11/10}} K^{\textrm{AdS}_{2}}_{\Delta_{1}-9/10}(\tau_{a},z_{a};\tau_{1})K^{\textrm{AdS}_{2}}_{\Delta_{2}-9/10}(\tau_{a},z_{a};\tau_{2}) \\
&\times G^{\textrm{AdS}_{2}}_{\Delta_{I}-9/10}(\tau_{a},z_{a};\tau_{b},z_{b})K^{\textrm{AdS}_{2}}_{\Delta_{3}-9/10}(\tau_{b},z_{b};\tau_{3})K^{\textrm{AdS}_{2}}_{\Delta_{4}-9/10}(\tau_{b},z_{b};\tau_{4}) \ .
\end{split}
\end{equation}
This integral is daunting because it involves a double integral over a hypergeometric function, but this is standard for Witten diagrams, even in maximally symmetric spacetimes. 

In order to evaluate higher point Witten diagrams, it is convenient to convert them to generalizations of Eq. (\ref{3pointfinal}). We first use a Schwinger-like representation of the bulk-to-bulk propagator:
\begin{equation}
\begin{split}
G^{\textrm{BFSS}}_{\Delta}(z_{1},\tau_{1};z_{2},\tau_{2})&\propto z_1^{\Delta } z_2^{\Delta } \int_{0}^{\infty} du_{1}du_{2}u_1^{\Delta -\frac{19}{10}} u_2^{\Delta -\frac{19}{10}} \left(u_1+u_2\right){}^{\frac{9}{10}-\Delta }   \\
&\times \exp \left(-\left(\left(u_1+u_2\right) \left(\tau_{12}^{2}+z_{12}^{2}\right)\right)-4 u_1 z_1 z_2\right)
\end{split}
\end{equation}
We then apply Eq. (\ref{swchingerrepre}) to all bulk-to-boundary propagators and evaluate all the bulk $\tau_{i}$ integrals, but not the $z_{i}$ integrals. The result is similar to a Feynman integral in Schwinger form. We can convert the result to an Euler integral using the same techniques as Section \ref{threepointcorrelator}, except we now treat the bulk $z_{i}$ variables as Schwinger parameters as well. At three-point, the bulk integral over $z$ could be evaluated without issue because the integrand was invariant under $z\rightarrow -z$. This is no longer true above three-point. The total number of Schwinger parameters is given by 
\begin{equation}
S=2B+V+n
\end{equation}
where $B$ is the number of bulk-to-bulk propagators, $V$ is the number of bulk vertices and $n$ is the number of bulk-to-boundary propagators. Given the large amount of combinatoric structure contained within the Schwinger representation of Feynman integrals, it is likely that this representation of Witten diagrams also encodes a lot of interesting structure.

\section{Discussion}

We considered the leading-order contributions to higher-point correlators in the BFSS matrix model. This paper is only an initial foray into a precision study of the BFSS matrix model in the holographic regime and there are many future directions. In the regime we studied, we only looked at the simplest non-trivial correlator. One could consider higher-point correlators, higher-loop corrections, and worldsheet corrections. Recent computational techniques developed to study higher point correlators in de Sitter and AdS, such as differential and difference equations \cite{Arkani-Hamed:2023kig,Alaverdian:2024llo}, should be immediately applicable. Certain three-point correlators might be protected by supersymmetry as in $\mathcal{N}=4$ super-Yang Mills \cite{Lee:1998bxa}.\footnote{We thank Henry Lin for discussion on this point.}

It is desirable to move away from the zero temperature limit, even though this makes the corresponding semi-classical gravity calculations much more difficult. Refs. \cite{Craps:2016cgo,Biggs:2023sqw} made progress in this direction by computing the quasi-normal modes. For example, one could consider the computation of the Euclidean two-point function at finite temperature. From a mathematical perspective, the two-point function simply corresponds to computing the Green's function of a second order differential equation. Alternatively, one could consider higher point functions in different approximations, such as in the limits of large dimension \cite{Fitzpatrick:2019zqz,Grinberg:2020fdj,Rodriguez-Gomez:2021pfh,Rodriguez-Gomez:2021mkk} or small temperature \cite{Bajc:2012vk,Bajc:2022wws}.  

One could consider the real-time dynamics of the BFSS matrix model, which can be derived via an analytic continuation of the Euclidean correlators. Even though real-time dynamics is difficult to access via Monte-Carlo due to the sign problem, a quantum simulation would face no such difficulties. One only needs roughly 7000 qubits to see black hole features \cite{Maldacena:2023acv}. A tantalizing possibility is being able to compare the real-time gravitational dynamics computed via numeric relativity\footnote{Such as black hole mergers.} with a quantum simulation of the BFSS model. 

\subsection*{Acknowledgments}

We thank Tom Banks, Henry Lin, Juan Maldacena, Sebastian Mizera, Giulio Salvatori, and Michael Winer for helpful discussion. AH is grateful to the Simons Foundation as well as the Edward and Kiyomi Baird Founders’ Circle Member Recognition for their support. We used Fable (Anthropic) and Sol (ChatGPT) to identify typographical and grammatical errors in version 2 of this manuscript.

\appendix

\section{Computing the action to cubic order}\label{quadraticcubicaction}

We consider the action in Eq. (\ref{action}). We outline the computation of the action to cubic order in this Appendix. We can truncate the spectrum to perturbations of the metric, the dilaton and the background gauge field at tree level. We only consider the perturbations about the metric given by Eq. (\ref{tensormodes}), and we set all other fields to be non-dynamical. In this approximation, the action simplifies dramatically. The last two terms in the action evaluate to constants:
\begin{equation}
\frac{16}{49}(\partial \phi)^{2}\rightarrow 9, \quad -\frac{1}{4}\left ( \frac{Nd_{0}}{(2\pi)^{2}}\right)^{-2}\left ( \frac{Nd_{0}}{(2\pi)^{2}} e^{\phi}\right )^{\frac{12}{7}}F_{2}^{2} \rightarrow \frac{49}{2}
\end{equation}
so we only need to consider perturbations of $\sqrt{g}$ and the Ricci scalar by the $b_{\ell,\vec{m}}$ modes. These can be computed using xPert \cite{Brizuela:2008ra}. Integration by parts identities and the fact that the Riemann tensor of the S${}_{8}$ sphere is given by 
\begin{equation}
R_{IJKL}=g_{IK}g_{JL}-g_{IL}g_{JK}
\end{equation}
are sufficient to simplify the quadratic and cubic action. We now need to consider the integral over the sphere. We use spherical harmonic identities such as
\begin{equation}
\delta_{\ell,\ell'}\delta_{\vec {m},\vec{m}'}=\int d\Omega Y^{IJ}_{\ell,\vec{m}}Y_{\ell',\vec{m}',IJ} \ ,
\end{equation}
to evaluate the integral over S$_{8}$ for the quadratic part of the action. For the cubic part of the action, more complex integrals appear involving triplets of spherical harmonics, such as 
\begin{equation}
\int d\Omega_{8} Y^{IJ}(\nabla_{I}Y^{KL})(\nabla_{J}Y_{KL}) \ .
\end{equation}
These integrals can be evaluated using the same computation strategy as Appendix B of Ref. \cite{Lee:1998bxa}.

\section{Introduction to method of regions}
\label{intromethodregions}

The method of regions is a computational technique for studying Feynman integrals in particular limits. The goal is to consider a series expansion of the integrand before integration. In general, it is easier to evaluate the integrals that appear in such a series expansion than to evaluate the original integral. However, series expanding an integrand does not commute with integration unless the resulting series, post-integration, is uniformly convergent. This is not true for the integrals we wish to study. However, the method of regions provides a systematic approach to such integrals. We will restrict ourselves to Euler integrals of the type
\begin{equation}
\int \left [\prod_{i} du_{i} u_{i}^{s_{i}}\right ]\left [\prod_{j} p_{j}^{-t_{j}}\right ]
\end{equation}
where the polynomials $p_{j}$ are Laurent polynomials of the $u_{i}$ and some fixed variable $y$. We consider Euler integrals in the limit that $y$ is parametrically small.  

The core idea behind the method of regions is that taking a series expansion of the integrand after taking a particular re-scaling of the integration variables is equivalent to evaluating the original integral on a distinct contour determined by the re-scaling.\footnote{By ``distinct contour," we are referring to a contour with different start and/or end points.} The contours associated with a given re-scaling are the ``regions" in ``method of regions." To be more concrete, we consider alternative scalings of the integration variables with $y$:
\begin{equation}\label{rscalelin}
u_{i}\rightarrow y^{\gamma_{i}}u_{i}
\end{equation}
where each scaling of the integration variables is denoted by $\vec{\gamma}=(\gamma_{1},\gamma_{2},\ldots)$. Each $\vec{\gamma}$ is associated with a distinct contour. To reproduce the original integrand/contour, we must sum multiple series expansions whose corresponding contours sum to the original contour. 

As an explanatory example, consider the following generalized Euler integral:
\begin{equation}\label{orgcalc}
\frac{2}{\sqrt{y}(1+\sqrt{y})}=\int_{0}^{\infty} du (1+u)^{-\frac{1}{2}}(1+yu)^{-\frac{3}{2}} \ .
\end{equation}
Suppose we simply series expand the integrand in $y$ using
\begin{equation}
(1+yu)^{-\frac{3}{2}}\rightarrow \sum_{i=0}^{\infty} u^i y^i \binom{-\frac{3}{2}}{i}
\end{equation}
integrate each term\footnote{Here and below, divergent term-by-term integrals are defined by analytic continuation in the exponents. In particular, although $\int_{0}^{\infty} du u^i(1+u)^{-\frac{1}{2}}$ diverges at its upper endpoint for $i=0,1,\ldots$, its analytically continued value is $\frac{\Gamma(i+1)\Gamma(-i-\frac{1}{2})}{\Gamma(\frac{1}{2})}$.} and then re-sum. The result is the contribution from the region with $u\sim \mathcal{O}(y^{0})$:
\begin{equation}\label{region1}
\textrm{Region }1: \quad -2\sum_{i=0}^{\infty}y^i=-\frac{2}{1-y} \ .
\end{equation}
Crucially, Eq. (\ref{region1}) is not equal to Eq. (\ref{orgcalc}). Instead, the integral representation of Eq. (\ref{region1}) is
\begin{equation}\label{region1contint}
-\frac{2}{1-y}=\int_{0}^{-1} du (1+u)^{-\frac{1}{2}}(1+yu)^{-\frac{3}{2}} \ .
\end{equation}
This shows that series expanding the given integrand in $y$ is equivalent to changing the contour from $u\in [0,\infty]$ to $u\in [0,-1]$. To reproduce the original contour, we need to consider another region. The procedure in Ref. \cite{Pak:2010pt} produces a second region associated with the re-scaling
\begin{equation}\label{region2den}
\textrm{Region }2\textrm{ rescaling}: \quad u\rightarrow u/y \ .
\end{equation}
To compute the contribution of the second region, we re-scale the integration variable according to Eq. (\ref{region2den}) and series expand in $y$. The result is the contribution from the region with $u\sim \mathcal{O}(y^{-1})$:
\begin{equation}\label{region2}
\textrm{Region }2: \quad 2y^{-\frac{1}{2}}\sum_{i=0}^{\infty}y^i=\frac{2}{\sqrt{y}(1-y)} \ .
\end{equation}
As expected, this second region contribution also can be identified with the integral in Eq. (\ref{orgcalc}), but evaluated on a distinct contour:
\begin{equation}\label{region2contint}
\frac{2}{\sqrt{y}(1-y)}=\int_{-1}^{\infty} du (1+u)^{-\frac{1}{2}}(1+yu)^{-\frac{3}{2}} \ .
\end{equation}
Upon summing the contributions of Eq. (\ref{region1contint}) and (\ref{region2contint}), we reproduce the original integral Eq. (\ref{orgcalc}). Indeed, $-\frac{2}{1-y}+\frac{2}{\sqrt{y}(1-y)}=\frac{2}{\sqrt{y}(1+\sqrt{y})}$, and the sum of the contours $u\in [0,-1]$ and $u\in[-1,\infty]$ is $u\in[0,\infty]$.

The method of regions is a collection of algorithms for finding all the required re-scalings whose associated contours sum to the original contour. This task can become somewhat subtle for generic integrals and the re-scalings in Eq. (\ref{rscalelin}) are often insufficient. For Feynman integrals, finding all the necessary regions is as difficult as finding all solutions to the Landau equations \cite{Jantzen:2011nz,Jantzen:2012mw}. There are different algorithms, such as the one in Ref. \cite{Semenova:2018cwy}, but efficiently finding all regions is still an open area of research. Fortunately, the algorithm in Ref. \cite{Pak:2010pt} is sufficient for the examples we study.

We will not derive the prescription in Ref. \cite{Pak:2010pt}, but simply state the procedure.\footnote{The intuition behind this algorithm is that we are computing the tropical version of the integral by tropicalizing the integrand and integration contour. See Ref. \cite{maclagan2015introduction} for a review of tropical geometry. Roughly speaking, tropical geometry corresponds to the study of functions at parametrically small/large values of their input parameters, so it is natural tropical geometry would be useful here.} First, one takes the product of all polynomials, $p_{i}$, and then computes the corresponding Newton polytope. The Newton polytope of a polynomial amounts to mapping each term in the polynomial to a vertex, where the location of the point is determined by the relevant exponents. For example, the polynomial
\begin{equation}
P=1+u_{1}^{2}+u_{2}^{2}+u_{1}^{2}u_{2}+u_{2}^{2}u_{1}+u_{2}^{3}u_{1}
\end{equation}
maps to the vertices
\begin{equation}
\textrm{Vertices}:\quad (0,0), \quad (2,0), \quad (0,2), \quad (2,1), \quad (1,2), \quad (1,3)
\end{equation}
where the $x$-axis corresponds to $u_{1}$ and the $y$-axis to $u_{2}$. For an Euler integral, the relevant variables are all the integration variables and the external variable we are taking to be small. The relevant regions correspond to all facets of the polytope  that point downward along the axis corresponding to the variable we are taking small.\footnote{By pointing downward, we mean the vector normal to the facet, orientated so it is pointing inward, is pointing in the upward direction along the relevant axis.} For each such facet, we compute the normal vector $\vec{f}_{i}$ with the convention that the component of $\vec{f}_{i}$ corresponding to the external variable is normalized to one. We then truncate $\vec{f}_{i}$ to the components corresponding to integration variables, $\vec{f}' \subset \vec{f} $ and identify $\vec{\gamma}_{i}=\vec{f}_{i}'$, which gives how the integration variables should scale in the corresponding region.  

\begin{figure}[t]
	\centering
	\begin{tikzpicture}[every node/.style={font=\footnotesize}]
	\filldraw [color=orange!30] (2,0) -- (0,0) -- (2,2) -- (4,2) -- (2,0);
	\draw [thick,-stealth] (0,0) -- (3.75,0) node[below=0pt]{$x$-axis ($u$)};
	\draw [thick,-stealth] (0,0) -- (0,3.75) node[left=0pt]{$y$-axis ($y$)};
    \draw [thick,>-stealth] (1,0) -- (1,0.5)node[right=0pt]{$\vec{\gamma}_{1}$};
    \draw [thick,>-stealth] (3,1) -- (3-0.5,1+0.5)node[below=0pt]{$\vec{\gamma}_{2}$};
	\draw [thick] (2,2) -- (4,2) -- (2,0) -- (0,0) -- (2,2);
	\filldraw (0,0) circle (1pt);
	\filldraw (2,2) circle (1pt) node[above=0pt]{$(1,1)$};
    \filldraw (2,0) circle (1pt) node[below=0pt]{$(1,0)$};
    \filldraw (4,2) circle (1pt) node[right=0pt]{$(2,1)$};
    \filldraw (0,0) circle (1pt) node[left=0pt]{$(0,0)$};
	\end{tikzpicture}
	\caption{The Newton polytope of the polynomial in Eq. (\ref{polystudy}). $u$ and $y$ correspond to the $x$ and $y$ axes respectively. $\vec{\gamma}_{1}=(0,1)$ and $\vec{\gamma}_{2}=(-1,1)$ are the normal vectors to the downward pointing facets.}
	\label{fig:semidirprod}
\end{figure}
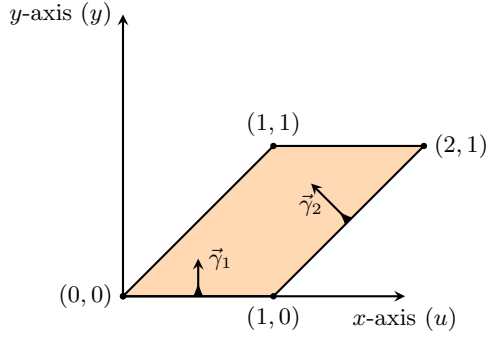
As an illustrative example, consider again the integral in Eq. (\ref{orgcalc}). We will derive the two region contributions from the corresponding Newton polytope. The Newton polytope of the relevant polynomial
\begin{equation}\label{polystudy}
p_{1}p_{2}=(1+u)(1+yu)=1+u+uy+yu^2
\end{equation}
is characterized by the vertices 
\begin{equation}
\textrm{Vertices}: \quad (0,0), \quad (1,0), \quad (1,1), \quad (2,1) \ .
\end{equation}
The $u$ variable and $y$ variable correspond to the $x$-axis and $y$-axis, see Fig.~\ref{fig:semidirprod}.  There are two downward facing facets:
\begin{equation}
\begin{split}
\textrm{Region 1 Facet}:& \quad \vec{\gamma}_{1}=(0,1) \\
\textrm{Region 2 Facet}:& \quad \vec{\gamma}_{2}=(-1,1) 
\end{split}
\end{equation}
which map onto the re-scalings 
\begin{equation}\label{regionsexf}
\begin{split}
\textrm{Region 1 Re-scaling}:& \quad u\rightarrow y^{0}u \\
\textrm{Region 2 Re-scaling}:& \quad u\rightarrow y^{-1}u \\
\end{split}
\end{equation}
These are the two regions identified previously, with contributions from each region given in Eqs. (\ref{region1}) and (\ref{region2}).

\section{Derivation of Eq. (\ref{eq:crucialiden2p})}\label{derivationequ}

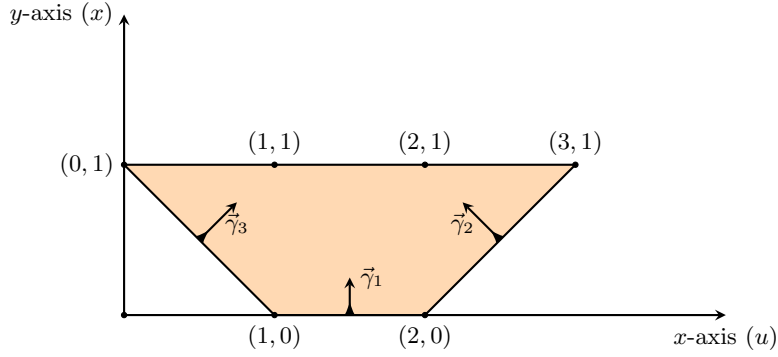
\begin{figure}[t]
	\centering
	\begin{tikzpicture}[every node/.style={font=\footnotesize}]
	\filldraw [color=orange!30] (2,0) -- (4,0) -- (6,2) -- (4,2) -- (2,2) -- (0,2) -- (2,0);
	\draw [thick,-stealth] (0,0) -- (8,0) node[below=0pt]{$x$-axis ($u$)};
	\draw [thick,-stealth] (0,0) -- (0,4) node[left=0pt]{$y$-axis ($x$)};
    \draw [thick,>-stealth] (3,0) -- (3,0.5)node[right=0pt]{$\vec{\gamma}_{1}$};
    \draw [thick,>-stealth] (5,1) -- (5-0.5,1+0.5)node[below=0pt]{$\vec{\gamma}_{3}$};
    \draw [thick,>-stealth] (1,1) -- (1+0.5,1+0.5)node[below=0pt]{$\vec{\gamma}_{2}$};
	\draw [thick] (2,0) -- (4,0) -- (6,2) -- (4,2) -- (2,2) -- (0,2) -- (2,0);
	\filldraw (0,0) circle (1pt);
	\filldraw (0,2) circle (1pt) node[left=0pt]{$(0,1)$};
    \filldraw (2,2) circle (1pt) node[above=0pt]{$(1,1)$};
    \filldraw (4,2) circle (1pt) node[above=0pt]{$(2,1)$};
    \filldraw (6,2) circle (1pt) node[above=0pt]{$(3,1)$};
    \filldraw (4,0) circle (1pt) node[below=0pt]{$(2,0)$};
    \filldraw (2,0) circle (1pt) node[below=0pt]{$(1,0)$};
	\end{tikzpicture}
	\caption{The Newton polytope of the vertices in Eq. (\ref{vertices3}). $u$ and $x$ correspond to the $x$ and $y$ axes respectively. $\vec{\gamma}_{1}=(0,1)$, $\vec{\gamma}_{2}=(1,1)$ and $\vec{\gamma}_{3}=(-1,1)$ are the normal vectors to the downward pointing facets.}
	\label{fig:newtonpoly3reg}
\end{figure}

We consider the derivation of Eq. (\ref{eq:crucialiden2p}) using method of regions. To simplify the problem, we consider integrals of the form 
\begin{equation}
f(\tau_{1}-\tau_{2},z_{b})=\int_{-\infty}^{\infty} d\tau \frac{1}{(z_{b}^{2}+(\tau-\tau_{1})^{2})^{A_{1}}}\frac{1}{(z_{b}^{2}+(\tau-\tau_{2})^{2})^{A_{2}}} \ .
\end{equation}
We can evaluate this integral using method of regions. We  first convert this integral to an Euler integral by applying
\begin{equation}
\frac{1}{[z^{2}+(\tau-\tau_{i})^{2}]^{A_{i}}}=\frac{1}{\Gamma(A_{i})}\int d\alpha_{i}\alpha_{i}^{A_{i}-1}e^{-\alpha_{i}[z^{2}+(\tau-\tau_{i})^{2}+i0]}
\end{equation}
and then integrating over $\tau$. We then insert 
\begin{equation}
1=\int d\gamma \delta(\gamma-\alpha_{1}-\alpha_{2})
\end{equation}
perform the rescaling $\alpha_{i}\rightarrow \gamma \alpha_{i}$, and integrate over $\gamma$. The result is 
\begin{equation}
\begin{split}
f(\tau_{1}-\tau_{2},z_{b})&=\frac{\sqrt{\pi}\Gamma \left(A_1+A_2-\frac{1}{2}\right) }{\Gamma \left(A_1\right) \Gamma \left(A_2\right)}\int d^{2}\alpha_{i} \alpha _1^{A_1-1} \alpha _2^{A_2-1} \left(\alpha _1+\alpha _2\right){}^{A_1+A_2-1} \\
&\times \left(\alpha _2 \alpha _1 \left(2 z_b^2+(\tau_{1}-\tau_{2})^2\right)+\alpha _1^2 z_b^2+\alpha _2^2 z_b^2\right){}^{-A_1-A_2+\frac{1}{2}} \delta(1-\alpha_{1}-\alpha_{2})
\end{split}
\end{equation}
To convert this integral to an Euler integral, we evaluate the $\alpha_{1}$ integral and then make the substitution 
\begin{equation}
\alpha_{2}=\frac{u}{1+u}
\end{equation}
so the $u$ is an integral from zero to infinity. The end result is the expression 
\begin{equation}
f(\tau_{1}-\tau_{2},z_{b})=\frac{\sqrt{\pi}\Gamma \left(A_1+A_2-\frac{1}{2}\right) }{\Gamma \left(A_1\right) \Gamma \left(A_2\right)}\frac{1}{(\tau_{1}-\tau_{2})^{2(A_{1}+A_{2})-1}}g\left (\frac{z_{b}^{2}}{(\tau_{1}-\tau_{2})^{2}}\right )
\end{equation}
where
\begin{equation}\label{gxintegral}
\begin{split}
g(x)=\int_{0}^{\infty} du u^{A_2-1} (u+1)^{A_1+A_2-1} \left((u+1)^2 x+u\right)^{-A_1-A_2+\frac{1}{2}} \ .
\end{split}
\end{equation}
We can then apply the computation strategy in Ref. \cite{Pak:2010pt} to find the integral in the limit that $z_{b}^{2}(\tau_{1}-\tau_{2})^{-2}=x<<1$. We compute the product of polynomials appearing in Eq. (\ref{gxintegral}) and find the corresponding Newton polytope, shown in Fig.~\ref{fig:newtonpoly3reg}, with vertices
\begin{equation}\label{vertices3}
\textrm{Vertices}:\quad (1,0), \quad (0,1), \quad (2,0),\quad (3,1), \quad (1,1), \quad (2,1) \ .
\end{equation}
There are three downward facing facets, which correspond to three regions 
\begin{equation}
\begin{split}
\textrm{Region 1}:& \quad u\sim \mathcal{O}(x^{0}) \\
\textrm{Region 2}:& \quad u\sim \mathcal{O}(x^{1}) \\
\textrm{Region 3}:& \quad u\sim \mathcal{O}(x^{-1}) \\
\end{split}
\end{equation}
with leading contributions 
\begin{equation}
\begin{split}
\textrm{Leading Contribution of Region 1}:& \quad g\sim x^{0}\frac{\Gamma \left(\frac{1}{2}-A_1\right) \Gamma \left(\frac{1}{2}-A_2\right)}{\Gamma \left(-A_1-A_2+1\right)} \\
\textrm{Leading Contribution of Region 2}:& \quad g\sim \frac{x^{\frac{1}{2}-A_1} \Gamma \left(A_1-\frac{1}{2}\right) \Gamma \left(A_2\right)}{\Gamma \left(A_1+A_2-\frac{1}{2}\right)}\\
\textrm{Leading Contribution of Region 3}:& \quad g\sim \frac{x^{\frac{1}{2}-A_2} \Gamma \left(A_1\right) \Gamma \left(A_2-\frac{1}{2}\right)}{\Gamma \left(A_1+A_2-\frac{1}{2}\right)} \\
\end{split}
\end{equation}
To compute the leading contribution at small $x$, we can discard Region 1 and focus on the leading contributions of Regions 2 and 3. The end result is the right hand side of Eq. (\ref{eq:crucialiden2p}) when we substitute in the possible values of $A_{1}$ and $A_{2}$. To recover Eq. (\ref{eq:crucialiden2p}), we split the bracket in Eq. (\ref{almostlaurent}) as $$(9-5\Delta)z_{b}^{2}+5\Delta(\tau-\tau_{2})^{2}=(9-10\Delta)z_{b}^{2}+5\Delta[z_{b}^{2}+(\tau-\tau_{2})^{2}]$$ and apply the above analysis twice, with 
\begin{equation}
\begin{split}
(A_{1},A_{2})&\to(\Delta-\frac{9}{10},\Delta+\frac{1}{10}) \\
(A_{1},A_{2})&\to(\Delta-\frac{9}{10},\Delta-\frac{9}{10})
\end{split}
\end{equation}
Both terms contribute at
order $z_{b}^{14/5-2\Delta}$ and their sum reproduces Eq. (\ref{eq:crucialiden2p}).

\bibliographystyle{apsrev4-1long}
\bibliography{GeneralBibliography.bib}

@article{Asano:2012zt,
    author = "Asano, Yuhma and Ishiki, Goro and Okada, Takashi and Shimasaki, Shinji",
    title = "{Exact results for perturbative partition functions of theories with SU(2|4) symmetry}",
    eprint = "1211.0364",
    archivePrefix = "arXiv",
    primaryClass = "hep-th",
    reportNumber = "KUNS-2422",
    doi = "10.1007/JHEP02(2013)148",
    journal = "JHEP",
    volume = "02",
    pages = "148",
    year = "2013"
}

@article{Lin:2013jra,
    author = "Lin, Ying-Hsuan and Shao, Shu-Heng and Wang, Yifan and Yin, Xi",
    title = "{A Low Temperature Expansion for Matrix Quantum Mechanics}",
    eprint = "1304.1593",
    archivePrefix = "arXiv",
    primaryClass = "hep-th",
    doi = "10.1007/JHEP05(2015)136",
    journal = "JHEP",
    volume = "05",
    pages = "136",
    year = "2015"
}

@article{Kanitscheider:2009as,
    author = "Kanitscheider, Ingmar and Skenderis, Kostas",
    title = "{Universal hydrodynamics of non-conformal branes}",
    eprint = "0901.1487",
    archivePrefix = "arXiv",
    primaryClass = "hep-th",
    reportNumber = "ITFA-2009-01, NSF-KITP-2009-02",
    doi = "10.1088/1126-6708/2009/04/062",
    journal = "JHEP",
    volume = "04",
    pages = "062",
    year = "2009"
}

@article{Susskind:1997cw,
    author = "Susskind, Leonard",
    title = "{Another conjecture about M(atrix) theory}",
    eprint = "hep-th/9704080",
    archivePrefix = "arXiv",
    reportNumber = "SU-ITP-97-11",
    month = "4",
    year = "1997"
}

@article{Skenderis:1998dq,
    author = "Skenderis, Kostas",
    editor = "Lust, D. and Otto, H. J.",
    title = "{Field theory limit of branes and gauged supergravities}",
    eprint = "hep-th/9903003",
    archivePrefix = "arXiv",
    reportNumber = "SPIN-1999-02",
    doi = "10.1002/(SICI)1521-3978(20001)48:1/3<205::AID-PROP205>3.0.CO;2-F",
    journal = "Fortsch. Phys.",
    volume = "48",
    pages = "205--208",
    year = "2000"
}

@article{Kanitscheider:2008kd,
    author = "Kanitscheider, Ingmar and Skenderis, Kostas and Taylor, Marika",
    title = "{Precision holography for non-conformal branes}",
    eprint = "0807.3324",
    archivePrefix = "arXiv",
    primaryClass = "hep-th",
    reportNumber = "ITFA-2008-25",
    doi = "10.1088/1126-6708/2008/09/094",
    journal = "JHEP",
    volume = "09",
    pages = "094",
    year = "2008"
}

@article{Boonstra:1998mp,
    author = "Boonstra, H. J. and Skenderis, K. and Townsend, P. K.",
    title = "{The domain wall / QFT correspondence}",
    eprint = "hep-th/9807137",
    archivePrefix = "arXiv",
    reportNumber = "KUL-TF-98-30",
    doi = "10.1088/1126-6708/1999/01/003",
    journal = "JHEP",
    volume = "01",
    pages = "003",
    year = "1999"
}

@article{Horowitz:1991cd,
    author = "Horowitz, Gary T. and Strominger, Andrew",
    title = "{Black strings and P-branes}",
    reportNumber = "UCSBTH-91-06",
    doi = "10.1016/0550-3213(91)90440-9",
    journal = "Nucl. Phys. B",
    volume = "360",
    pages = "197--209",
    year = "1991"
}

@article{Berkowitz:2018qhn,
    author = "Berkowitz, Evan and Hanada, Masanori and Rinaldi, Enrico and Vranas, Pavlos",
    title = "{Gauged And Ungauged: A Nonperturbative Test}",
    eprint = "1802.02985",
    archivePrefix = "arXiv",
    primaryClass = "hep-th",
    reportNumber = "YITP-18-01, RBRC-1267",
    doi = "10.1007/JHEP06(2018)124",
    journal = "JHEP",
    volume = "06",
    pages = "124",
    year = "2018"
}

@article{Banks:1996vh,
    author = "Banks, Tom and Fischler, W. and Shenker, S. H. and Susskind, Leonard",
    title = "{M theory as a matrix model: A Conjecture}",
    eprint = "hep-th/9610043",
    archivePrefix = "arXiv",
    reportNumber = "RU-96-95, SU-ITP-96-12, UTTG-13-96",
    doi = "10.1103/PhysRevD.55.5112",
    journal = "Phys. Rev. D",
    volume = "55",
    pages = "5112--5128",
    year = "1997"
}

@article{Itzhaki:1998dd,
    author = "Itzhaki, Nissan and Maldacena, Juan Martin and Sonnenschein, Jacob and Yankielowicz, Shimon",
    title = "{Supergravity and the large N limit of theories with sixteen supercharges}",
    eprint = "hep-th/9802042",
    archivePrefix = "arXiv",
    reportNumber = "TAUP-2474-98, HUTP-98-A003",
    doi = "10.1103/PhysRevD.58.046004",
    journal = "Phys. Rev. D",
    volume = "58",
    pages = "046004",
    year = "1998"
}

@article{Berkowitz:2016jlq,
    author = "Berkowitz, Evan and Rinaldi, Enrico and Hanada, Masanori and Ishiki, Goro and Shimasaki, Shinji and Vranas, Pavlos",
    title = "{Precision lattice test of the gauge/gravity duality at large-$N$}",
    eprint = "1606.04951",
    archivePrefix = "arXiv",
    primaryClass = "hep-lat",
    reportNumber = "LLNL-JRNL-694385, UTHEP-689, YITP-16-73",
    doi = "10.1103/PhysRevD.94.094501",
    journal = "Phys. Rev. D",
    volume = "94",
    number = "9",
    pages = "094501",
    year = "2016"
}

@article{Pateloudis:2022ijr,
    author = {Pateloudis, Stratos and Bergner, Georg and Hanada, Masanori and Rinaldi, Enrico and Sch\"afer, Andreas and Vranas, Pavlos and Watanabe, Hiromasa and Bodendorfer, Norbert},
    title = "{Precision test of gauge/gravity duality in D0-brane matrix model at low temperature}",
    eprint = "2210.04881",
    archivePrefix = "arXiv",
    primaryClass = "hep-th",
    month = "10",
    year = "2022"
}

@article{Kitaev:2017awl,
    author = "Kitaev, Alexei and Suh, S. Josephine",
    title = "{The soft mode in the Sachdev-Ye-Kitaev model and its gravity dual}",
    eprint = "1711.08467",
    archivePrefix = "arXiv",
    primaryClass = "hep-th",
    doi = "10.1007/JHEP05(2018)183",
    journal = "JHEP",
    volume = "05",
    pages = "183",
    year = "2018"
}

@article{Maldacena:2016hyu,
    author = "Maldacena, Juan and Stanford, Douglas",
    title = "{Remarks on the Sachdev-Ye-Kitaev model}",
    eprint = "1604.07818",
    archivePrefix = "arXiv",
    primaryClass = "hep-th",
    doi = "10.1103/PhysRevD.94.106002",
    journal = "Phys. Rev. D",
    volume = "94",
    number = "10",
    pages = "106002",
    year = "2016"
}

@article{Kabat:1997sa,
    author = "Kabat, Daniel N. and Taylor, Washington",
    title = "{Linearized supergravity from matrix theory}",
    eprint = "hep-th/9712185",
    archivePrefix = "arXiv",
    reportNumber = "IASSNS-HEP-97-141, PUPT-1751",
    doi = "10.1016/S0370-2693(98)00281-0",
    journal = "Phys. Lett. B",
    volume = "426",
    pages = "297--305",
    year = "1998"
}

@article{Taylor:1998tv,
    author = "Taylor, Washington and Van Raamsdonk, Mark",
    title = "{Supergravity currents and linearized interactions for matrix theory configurations with fermionic backgrounds}",
    eprint = "hep-th/9812239",
    archivePrefix = "arXiv",
    reportNumber = "PUPT-1828, MIT-CTP-2810",
    doi = "10.1088/1126-6708/1999/04/013",
    journal = "JHEP",
    volume = "04",
    pages = "013",
    year = "1999"
}

@article{deWit:1988wri,
    author = "de Wit, B. and Hoppe, J. and Nicolai, H.",
    title = "{On the Quantum Mechanics of Supermembranes}",
    reportNumber = "THU-88-15, KA-THEP-6/88",
    doi = "10.1016/0550-3213(88)90116-2",
    journal = "Nucl. Phys. B",
    volume = "305",
    pages = "545",
    year = "1988"
}

@article{Seiberg:1997ad,
    author = "Seiberg, Nathan",
    title = "{Why is the matrix model correct?}",
    eprint = "hep-th/9710009",
    archivePrefix = "arXiv",
    reportNumber = "IASSNS-HEP-97-108",
    doi = "10.1103/PhysRevLett.79.3577",
    journal = "Phys. Rev. Lett.",
    volume = "79",
    pages = "3577--3580",
    year = "1997"
}

@article{Sen:1997we,
    author = "Sen, Ashoke",
    title = "{D0-branes on T**n and matrix theory}",
    eprint = "hep-th/9709220",
    archivePrefix = "arXiv",
    reportNumber = "MRI-PHY-P97-0926",
    doi = "10.4310/ATMP.1998.v2.n1.a2",
    journal = "Adv. Theor. Math. Phys.",
    volume = "2",
    pages = "51--59",
    year = "1998"
}

@article{Polchinski:1999br,
    author = "Polchinski, Joseph",
    editor = "Iso, S. and Kawai, H. and Natsuume, M.",
    title = "{M theory and the light cone}",
    eprint = "hep-th/9903165",
    archivePrefix = "arXiv",
    reportNumber = "NSF-ITP-99-17",
    doi = "10.1143/PTPS.134.158",
    journal = "Prog. Theor. Phys. Suppl.",
    volume = "134",
    pages = "158--170",
    year = "1999"
}

@article{Miller:2022fvc,
    author = "Miller, Noah and Strominger, Andrew and Tropper, Adam and Wang, Tianli",
    title = "{Soft gravitons in the BFSS matrix model}",
    eprint = "2208.14547",
    archivePrefix = "arXiv",
    primaryClass = "hep-th",
    doi = "10.1007/JHEP11(2023)174",
    journal = "JHEP",
    volume = "11",
    pages = "174",
    year = "2023"
}

@article{Tropper:2023fjr,
    author = "Tropper, Adam and Wang, Tianli",
    title = "{Lorentz symmetry and IR structure of the BFSS matrix model}",
    eprint = "2303.14200",
    archivePrefix = "arXiv",
    primaryClass = "hep-th",
    doi = "10.1007/JHEP07(2023)150",
    journal = "JHEP",
    volume = "07",
    pages = "150",
    year = "2023"
}

@article{Herderschee:2023bnc,
    author = "Herderschee, Aidan and Maldacena, Juan",
    title = "{Soft Theorems in Matrix Theory}",
    eprint = "2312.15111",
    archivePrefix = "arXiv",
    primaryClass = "hep-th",
    month = "12",
    year = "2023"
}

@article{Bergner:2021goh,
    author = {Bergner, Georg and Bodendorfer, Norbert and Hanada, Masanori and Pateloudis, Stratos and Rinaldi, Enrico and Sch\"afer, Andreas and Vranas, Pavlos and Watanabe, Hiromasa},
    collaboration = "Monte Carlo String/M-theory (MCSMC), MCSMC",
    title = "{Confinement/deconfinement transition in the D0-brane matrix model \textemdash{} A signature of M-theory?}",
    eprint = "2110.01312",
    archivePrefix = "arXiv",
    primaryClass = "hep-th",
    reportNumber = "LLNL-JRNL-824792, RIKEN-iTHEMS-Report-21, UTHEP-759, DMUS-MP-21/13",
    doi = "10.1007/JHEP05(2022)096",
    journal = "JHEP",
    volume = "05",
    pages = "096",
    year = "2022"
}

@article{Freedman:1998tz,
    author = "Freedman, Daniel Z. and Mathur, Samir D. and Matusis, Alec and Rastelli, Leonardo",
    title = "{Correlation functions in the CFT(d) / AdS(d+1) correspondence}",
    eprint = "hep-th/9804058",
    archivePrefix = "arXiv",
    reportNumber = "MIT-CTP-2727",
    doi = "10.1016/S0550-3213(99)00053-X",
    journal = "Nucl. Phys. B",
    volume = "546",
    pages = "96--118",
    year = "1999"
}

@article{DHoker:1999mqo,
    author = "D'Hoker, Eric and Freedman, Daniel Z. and Rastelli, Leonardo",
    title = "{AdS / CFT four point functions: How to succeed at z integrals without really trying}",
    eprint = "hep-th/9905049",
    archivePrefix = "arXiv",
    reportNumber = "MIT-CTP-2857, UCLA-99-TEP-16",
    doi = "10.1016/S0550-3213(99)00526-X",
    journal = "Nucl. Phys. B",
    volume = "562",
    pages = "395--411",
    year = "1999"
}

@article{Maldacena:1997re,
    author = "Maldacena, Juan Martin",
    title = "{The Large N limit of superconformal field theories and supergravity}",
    eprint = "hep-th/9711200",
    archivePrefix = "arXiv",
    reportNumber = "HUTP-97-A097, HUTP-98-A097",
    doi = "10.4310/ATMP.1998.v2.n2.a1",
    journal = "Adv. Theor. Math. Phys.",
    volume = "2",
    pages = "231--252",
    year = "1998"
}

@article{Witten:1998qj,
    author = "Witten, Edward",
    title = "{Anti-de Sitter space and holography}",
    eprint = "hep-th/9802150",
    archivePrefix = "arXiv",
    reportNumber = "IASSNS-HEP-98-15",
    doi = "10.4310/ATMP.1998.v2.n2.a2",
    journal = "Adv. Theor. Math. Phys.",
    volume = "2",
    pages = "253--291",
    year = "1998"
}

@article{Gubser:1998bc,
    author = "Gubser, S. S. and Klebanov, Igor R. and Polyakov, Alexander M.",
    title = "{Gauge theory correlators from noncritical string theory}",
    eprint = "hep-th/9802109",
    archivePrefix = "arXiv",
    reportNumber = "PUPT-1767",
    doi = "10.1016/S0370-2693(98)00377-3",
    journal = "Phys. Lett. B",
    volume = "428",
    pages = "105--114",
    year = "1998"
}

@article{aomoto1975equations,
  title={Les {\'e}quations aux diff{\'e}rences lin{\'e}aires et les int{\'e}grales des fonctions multiformes},
  author={Aomoto, Kazuhiko},
  journal={J. Fac. Sci. Univ. Tokyo},
  volume={22},
  number={3},
  pages={271--297},
  year={1975}
}

@inproceedings{gel1986general,
  title={General theory of hypergeometric functions},
  author={Gel'fand, Izrail Moiseevich},
  booktitle={Doklady Akademii Nauk},
  volume={288},
  number={1},
  pages={14--18},
  year={1986},
  organization={Russian Academy of Sciences}
}

@article{Gelfand:1990bua,
    author = "Gelfand, I. M and Kapranov, M. M and Zelevinsky, A. V",
    title = "{Generalized Euler integrals and 
				A-hypergeometric functions
			}",
    doi = "10.1016/0001-8708(90)90048-R",
    journal = "Adv. Math.",
    volume = "84",
    number = "2",
    pages = "255--271",
    year = "1990"
}

@article{Matsubara-Heo:2023ylc,
    author = "Matsubara-Heo, Saiei-Jaeyeong and Mizera, Sebastian and Telen, Simon",
    title = "{Four lectures on Euler integrals}",
    eprint = "2306.13578",
    archivePrefix = "arXiv",
    primaryClass = "math-ph",
    doi = "10.21468/SciPostPhysLectNotes.75",
    journal = "SciPost Phys. Lect. Notes",
    volume = "75",
    pages = "1",
    year = "2023"
}

@book{Smirnov:1991jn,
    author = "Smirnov, V. A.",
    title = "{Renormalization and asymptotic expansions}",
    volume = "14",
    year = "1991"
}

@article{Beneke:1997zp,
    author = "Beneke, M. and Smirnov, Vladimir A.",
    title = "{Asymptotic expansion of Feynman integrals near threshold}",
    eprint = "hep-ph/9711391",
    archivePrefix = "arXiv",
    reportNumber = "CERN-TH-97-315",
    doi = "10.1016/S0550-3213(98)00138-2",
    journal = "Nucl. Phys. B",
    volume = "522",
    pages = "321--344",
    year = "1998"
}

@article{Smirnov:1998vk,
    author = "Smirnov, Vladimir A. and Rakhmetov, E. R.",
    title = "{The Strategy of regions for asymptotic expansion of two loop vertex Feynman diagrams}",
    eprint = "hep-ph/9812529",
    archivePrefix = "arXiv",
    reportNumber = "NPI-MSU-98-54-555",
    doi = "10.1007/BF02557396",
    journal = "Theor. Math. Phys.",
    volume = "120",
    pages = "870--875",
    year = "1999"
}

@article{Smirnov:1999bza,
    author = "Smirnov, Vladimir A.",
    title = "{Problems of the strategy of regions}",
    eprint = "hep-ph/9907471",
    archivePrefix = "arXiv",
    doi = "10.1016/S0370-2693(99)01061-8",
    journal = "Phys. Lett. B",
    volume = "465",
    pages = "226--234",
    year = "1999"
}

@article{Pak:2010pt,
    author = "Pak, A. and Smirnov, A.",
    title = "{Geometric approach to asymptotic expansion of Feynman integrals}",
    eprint = "1011.4863",
    archivePrefix = "arXiv",
    primaryClass = "hep-ph",
    doi = "10.1140/epjc/s10052-011-1626-1",
    journal = "Eur. Phys. J. C",
    volume = "71",
    pages = "1626",
    year = "2011"
}

@article{Ananthanarayan:2018tog,
    author = "Ananthanarayan, B. and Pal, Abhishek and Ramanan, S. and Sarkar, Ratan",
    title = "{Unveiling Regions in multi-scale Feynman Integrals using Singularities and Power Geometry}",
    eprint = "1810.06270",
    archivePrefix = "arXiv",
    primaryClass = "hep-ph",
    doi = "10.1140/epjc/s10052-019-6533-x",
    journal = "Eur. Phys. J. C",
    volume = "79",
    number = "1",
    pages = "57",
    year = "2019"
}

@article{Ananthanarayan:2020ptw,
    author = "Ananthanarayan, B. and Das, Abhijit B. and Sarkar, Ratan",
    title = "{Asymptotic analysis of Feynman diagrams and their maximal cuts}",
    eprint = "2003.02451",
    archivePrefix = "arXiv",
    primaryClass = "hep-ph",
    doi = "10.1140/epjc/s10052-020-08609-0",
    journal = "Eur. Phys. J. C",
    volume = "80",
    number = "12",
    pages = "1131",
    year = "2020"
}

@article{Jantzen:2011nz,
    author = "Jantzen, Bernd",
    title = "{Foundation and generalization of the expansion by regions}",
    eprint = "1111.2589",
    archivePrefix = "arXiv",
    primaryClass = "hep-ph",
    reportNumber = "TTK-11-53, SFB-CPP-11-61",
    doi = "10.1007/JHEP12(2011)076",
    journal = "JHEP",
    volume = "12",
    pages = "076",
    year = "2011"
}

@article{Jantzen:2012mw,
    author = "Jantzen, Bernd and Smirnov, Alexander V. and Smirnov, Vladimir A.",
    title = "{Expansion by regions: revealing potential and Glauber regions automatically}",
    eprint = "1206.0546",
    archivePrefix = "arXiv",
    primaryClass = "hep-ph",
    reportNumber = "TTK-12-21, TTP12-015, SFB-CPP-12-31",
    doi = "10.1140/epjc/s10052-012-2139-2",
    journal = "Eur. Phys. J. C",
    volume = "72",
    pages = "2139",
    year = "2012"
}

@article{Semenova:2018cwy,
    author = "Semenova, Tatiana Yu and Smirnov, Alexander V. and Smirnov, Vladimir A.",
    title = "{On the status of expansion by regions}",
    eprint = "1809.04325",
    archivePrefix = "arXiv",
    primaryClass = "hep-th",
    doi = "10.1140/epjc/s10052-019-6653-3",
    journal = "Eur. Phys. J. C",
    volume = "79",
    number = "2",
    pages = "136",
    year = "2019"
}

@article{Sekino:1999av,
    author = "Sekino, Yasuhiro and Yoneya, Tamiaki",
    title = "{Generalized AdS / CFT correspondence for matrix theory in the large N limit}",
    eprint = "hep-th/9907029",
    archivePrefix = "arXiv",
    reportNumber = "UT-KOMABA-99-8",
    doi = "10.1016/S0550-3213(99)00793-2",
    journal = "Nucl. Phys. B",
    volume = "570",
    pages = "174--206",
    year = "2000"
}

@article{Biggs:2023sqw,
    author = "Biggs, Anna and Maldacena, Juan",
    title = "{Scaling similarities and quasinormal modes of D0 black hole solutions}",
    eprint = "2303.09974",
    archivePrefix = "arXiv",
    primaryClass = "hep-th",
    doi = "10.1007/JHEP11(2023)155",
    journal = "JHEP",
    volume = "11",
    pages = "155",
    year = "2023"
}

@article{Maldacena:2023acv,
    author = "Maldacena, Juan",
    title = "{A simple quantum system that describes a black hole}",
    eprint = "2303.11534",
    archivePrefix = "arXiv",
    primaryClass = "hep-th",
    month = "3",
    year = "2023"
}

@article{Grinberg:2020fdj,
    author = "Grinberg, Matan and Maldacena, Juan",
    title = "{Proper time to the black hole singularity from thermal one-point functions}",
    eprint = "2011.01004",
    archivePrefix = "arXiv",
    primaryClass = "hep-th",
    doi = "10.1007/JHEP03(2021)131",
    journal = "JHEP",
    volume = "03",
    pages = "131",
    year = "2021"
}

@article{Fitzpatrick:2019zqz,
    author = "Fitzpatrick, A. Liam and Huang, Kuo-Wei",
    title = "{Universal Lowest-Twist in CFTs from Holography}",
    eprint = "1903.05306",
    archivePrefix = "arXiv",
    primaryClass = "hep-th",
    doi = "10.1007/JHEP08(2019)138",
    journal = "JHEP",
    volume = "08",
    pages = "138",
    year = "2019"
}

@article{Rodriguez-Gomez:2021pfh,
    author = "Rodriguez-Gomez, D. and Russo, J. G.",
    title = "{Correlation functions in finite temperature CFT and black hole singularities}",
    eprint = "2102.11891",
    archivePrefix = "arXiv",
    primaryClass = "hep-th",
    reportNumber = "ICCUB-21-002",
    doi = "10.1007/JHEP06(2021)048",
    journal = "JHEP",
    volume = "06",
    pages = "048",
    year = "2021"
}

@article{Rodriguez-Gomez:2021mkk,
    author = "Rodriguez-Gomez, D. and Russo, J. G.",
    title = "{Thermal correlation functions in CFT and factorization}",
    eprint = "2105.13909",
    archivePrefix = "arXiv",
    primaryClass = "hep-th",
    reportNumber = "ICCUB-21-007",
    doi = "10.1007/JHEP11(2021)049",
    journal = "JHEP",
    volume = "11",
    pages = "049",
    year = "2021"
}

@article{Bajc:2012vk,
    author = "Bajc, Borut and Lugo, Adrian R. and Sturla, Mauricio B.",
    title = "{Spontaneous breaking of a discrete symmetry and holography}",
    eprint = "1203.2636",
    archivePrefix = "arXiv",
    primaryClass = "hep-th",
    doi = "10.1007/JHEP04(2012)119",
    journal = "JHEP",
    volume = "04",
    pages = "119",
    year = "2012"
}

@article{Bajc:2022wws,
    author = "Bajc, Borut and Lugo, Adrian R.",
    title = "{Holographic thermal propagator for arbitrary scale dimensions}",
    eprint = "2212.13639",
    archivePrefix = "arXiv",
    primaryClass = "hep-th",
    doi = "10.1007/JHEP05(2023)103",
    journal = "JHEP",
    volume = "05",
    pages = "103",
    year = "2023"
}

@article{Craps:2016cgo,
    author = "Craps, Ben and Evnin, Oleg and Nguyen, K\'evin",
    title = "{Matrix Thermalization}",
    eprint = "1610.05333",
    archivePrefix = "arXiv",
    primaryClass = "hep-th",
    doi = "10.1007/JHEP02(2017)041",
    journal = "JHEP",
    volume = "02",
    pages = "041",
    year = "2017"
}

@article{Hanada:2009ne,
    author = "Hanada, Masanori and Nishimura, Jun and Sekino, Yasuhiro and Yoneya, Tamiaki",
    title = "{Monte Carlo studies of Matrix theory correlation functions}",
    eprint = "0911.1623",
    archivePrefix = "arXiv",
    primaryClass = "hep-th",
    reportNumber = "WS-14-09-NOV-DPPA, KEK-TH-1337, OIQP-09-12, UT-KOMABA-09-5",
    doi = "10.1103/PhysRevLett.104.151601",
    journal = "Phys. Rev. Lett.",
    volume = "104",
    pages = "151601",
    year = "2010"
}

@article{Hanada:2011fq,
    author = "Hanada, Masanori and Nishimura, Jun and Sekino, Yasuhiro and Yoneya, Tamiaki",
    title = "{Direct test of the gauge-gravity correspondence for Matrix theory correlation functions}",
    eprint = "1108.5153",
    archivePrefix = "arXiv",
    primaryClass = "hep-th",
    reportNumber = "KEK-TH-1490, KEK\_TH-1490",
    doi = "10.1007/JHEP12(2011)020",
    journal = "JHEP",
    volume = "12",
    pages = "020",
    year = "2011"
}

@article{Arkani-Hamed:2023kig,
    author = "Arkani-Hamed, Nima and Baumann, Daniel and Hillman, Aaron and Joyce, Austin and Lee, Hayden and Pimentel, Guilherme L.",
    title = "{Differential Equations for Cosmological Correlators}",
    eprint = "2312.05303",
    archivePrefix = "arXiv",
    primaryClass = "hep-th",
    month = "12",
    year = "2023"
}

@article{Diwakar:2021juk,
    author = "Diwakar, Pranav and Herderschee, Aidan and Roiban, Radu and Teng, Fei",
    title = "{BCJ amplitude relations for Anti-de Sitter boundary correlators in embedding space}",
    eprint = "2106.10822",
    archivePrefix = "arXiv",
    primaryClass = "hep-th",
    reportNumber = "LCTP-21-15",
    doi = "10.1007/JHEP10(2021)141",
    journal = "JHEP",
    volume = "10",
    pages = "141",
    year = "2021"
}

@article{Polchinski:2016xgd,
    author = "Polchinski, Joseph and Rosenhaus, Vladimir",
    title = "{The Spectrum in the Sachdev-Ye-Kitaev Model}",
    eprint = "1601.06768",
    archivePrefix = "arXiv",
    primaryClass = "hep-th",
    doi = "10.1007/JHEP04(2016)001",
    journal = "JHEP",
    volume = "04",
    pages = "001",
    year = "2016"
}

@article{Herderschee:2023pza,
    author = "Herderschee, Aidan and Maldacena, Juan",
    title = "{Three point amplitudes in matrix theory}",
    eprint = "2312.12592",
    archivePrefix = "arXiv",
    primaryClass = "hep-th",
    doi = "10.1088/1751-8121/ad389b",
    journal = "J. Phys. A",
    volume = "57",
    number = "16",
    pages = "165401",
    year = "2024"
}

@article{Hillman:2023ezp,
    author = "Hillman, Aaron",
    title = "{A Subtraction Scheme for Feynman Integrals}",
    eprint = "2311.03439",
    archivePrefix = "arXiv",
    primaryClass = "hep-th",
    month = "11",
    year = "2023"
}

@book{maclagan2015introduction,
  title={Introduction to Tropical Geometry},
  author={Maclagan, Diane and Sturmfels, Bernd},
  series={Graduate Studies in Mathematics},
  volume={161},
  year={2015},
  publisher={American Mathematical Society},
  isbn={978-0-8218-5198-2}
}

@article{DHoker:1998bqu,
    author = "D'Hoker, Eric and Freedman, Daniel Z.",
    title = "{Gauge boson exchange in AdS(d+1)}",
    eprint = "hep-th/9809179",
    archivePrefix = "arXiv",
    reportNumber = "UCLA-98-TEP-33, MIT-CTP-2781",
    doi = "10.1016/S0550-3213(98)00852-9",
    journal = "Nucl. Phys. B",
    volume = "544",
    pages = "612--632",
    year = "1999"
}

@article{DHoker:1998ecp,
    author = "D'Hoker, Eric and Freedman, Daniel Z.",
    title = "{General scalar exchange in AdS(d+1)}",
    eprint = "hep-th/9811257",
    archivePrefix = "arXiv",
    reportNumber = "UCLA-98-TEP-34, MIT-CTP-2795",
    doi = "10.1016/S0550-3213(99)00169-8",
    journal = "Nucl. Phys. B",
    volume = "550",
    pages = "261--288",
    year = "1999"
}

@article{Liu:1998th,
    author = "Liu, Hong",
    title = "{Scattering in anti-de Sitter space and operator product expansion}",
    eprint = "hep-th/9811152",
    archivePrefix = "arXiv",
    reportNumber = "IMPERIAL-TP-98-99-012",
    doi = "10.1103/PhysRevD.60.106005",
    journal = "Phys. Rev. D",
    volume = "60",
    pages = "106005",
    year = "1999"
}

@article{DHoker:1999bve,
    author = "D'Hoker, Eric and Freedman, Daniel Z. and Mathur, Samir D. and Matusis, Alec and Rastelli, Leonardo",
    title = "{Graviton and gauge boson propagators in AdS(d+1)}",
    eprint = "hep-th/9902042",
    archivePrefix = "arXiv",
    reportNumber = "MIT-CTP-2824, UCLA-99-TEP-1",
    doi = "10.1016/S0550-3213(99)00524-6",
    journal = "Nucl. Phys. B",
    volume = "562",
    pages = "330--352",
    year = "1999"
}

@article{DHoker:1999kzh,
    author = "D'Hoker, Eric and Freedman, Daniel Z. and Mathur, Samir D. and Matusis, Alec and Rastelli, Leonardo",
    title = "{Graviton exchange and complete four point functions in the AdS / CFT correspondence}",
    eprint = "hep-th/9903196",
    archivePrefix = "arXiv",
    reportNumber = "MIT-CTP-2843, UCLA-99-TEP-2",
    doi = "10.1016/S0550-3213(99)00525-8",
    journal = "Nucl. Phys. B",
    volume = "562",
    pages = "353--394",
    year = "1999"
}

@article{Alaverdian:2024llo,
    author = "Alaverdian, Mark and Herderschee, Aidan and Roiban, Radu and Teng, Fei",
    title = "{Difference equations and integral families for Witten diagrams}",
    eprint = "2406.04186",
    archivePrefix = "arXiv",
    primaryClass = "hep-th",
    doi = "10.1007/JHEP12(2024)070",
    journal = "JHEP",
    volume = "12",
    pages = "070",
    year = "2024"
}

@article{Brizuela:2008ra,
    author = "Brizuela, David and Martin-Garcia, Jose M. and Mena Marugan, Guillermo A.",
    title = "{xPert: Computer algebra for metric perturbation theory}",
    eprint = "0807.0824",
    archivePrefix = "arXiv",
    primaryClass = "gr-qc",
    doi = "10.1007/s10714-009-0773-2",
    journal = "Gen. Rel. Grav.",
    volume = "41",
    pages = "2415--2431",
    year = "2009"
}

@article{Lee:1998bxa,
    author = "Lee, Sangmin and Minwalla, Shiraz and Rangamani, Mukund and Seiberg, Nathan",
    title = "{Three point functions of chiral operators in D = 4, N=4 SYM at large N}",
    eprint = "hep-th/9806074",
    archivePrefix = "arXiv",
    reportNumber = "PUPT-1796, IASSNS-HEP-98-51, PUPT-1796-",
    doi = "10.4310/ATMP.1998.v2.n4.a1",
    journal = "Adv. Theor. Math. Phys.",
    volume = "2",
    pages = "697--718",
    year = "1998"
}

@article{Batra:2025ivy,
    author = "Batra, Gauri and Lin, Henry",
    title = "{Giant gravitons in D$p$-brane holography}",
    eprint = "2502.14249",
    archivePrefix = "arXiv",
    primaryClass = "hep-th",
    month = "2",
    year = "2025"
}

@mastersthesis{Penedones:2007ns,
    author = "Penedones, Joao",
    title = "{High Energy Scattering in the AdS/CFT Correspondence}",
    eprint = "0712.0802",
    archivePrefix = "arXiv",
    primaryClass = "hep-th",
    type = "Other thesis",
    month = "12",
    year = "2007"
}

@article{Cornalba:2007fs,
    author = "Cornalba, Lorenzo",
    title = "{Eikonal methods in AdS/CFT: Regge theory and multi-reggeon exchange}",
    eprint = "0710.5480",
    archivePrefix = "arXiv",
    primaryClass = "hep-th",
    month = "10",
    year = "2007"
}

@article{Cornalba:2008qf,
    author = "Cornalba, Lorenzo and Costa, Miguel S. and Penedones, Joao",
    title = "{Eikonal Methods in AdS/CFT: BFKL Pomeron at Weak Coupling}",
    eprint = "0801.3002",
    archivePrefix = "arXiv",
    primaryClass = "hep-th",
    reportNumber = "BICOCCA-FT-08-02, NSF-KITP-07-212",
    doi = "10.1088/1126-6708/2008/06/048",
    journal = "JHEP",
    volume = "06",
    pages = "048",
    year = "2008"
}

@article{Asano:2014vba,
    author = "Asano, Yuhma and Ishiki, Goro and Okada, Takashi and Shimasaki, Shinji",
    title = "{Emergent bubbling geometries in the plane wave matrix model}",
    eprint = "1401.5079",
    archivePrefix = "arXiv",
    primaryClass = "hep-th",
    doi = "10.1007/JHEP05(2014)075",
    journal = "JHEP",
    volume = "05",
    pages = "075",
    year = "2014"
}

@article{Asano:2017xiy,
    author = "Asano, Yuhma and Ishiki, Goro and Shimasaki, Shinji and Terashima, Seiji",
    title = "{Spherical transverse M5-branes in matrix theory}",
    eprint = "1701.07140",
    archivePrefix = "arXiv",
    primaryClass = "hep-th",
    reportNumber = "DIAS-STP-16-14, UTHEP-701, YITP-16-149",
    doi = "10.1103/PhysRevD.96.126003",
    journal = "Phys. Rev. D",
    volume = "96",
    number = "12",
    pages = "126003",
    year = "2017"
}

@inproceedings{Klebanov:2000me,
    author = "Klebanov, Igor R.",
    title = "{TASI lectures: Introduction to the AdS / CFT correspondence}",
    booktitle = "{Theoretical Advanced Study Institute in Elementary Particle Physics (TASI 99): Strings, Branes, and Gravity}",
    eprint = "hep-th/0009139",
    archivePrefix = "arXiv",
    reportNumber = "PUPT-1954",
    doi = "10.1142/9789812799630_0007",
    pages = "615--650",
    month = "9",
    year = "2000"
}

@article{Sekino:2000mg,
    author = "Sekino, Yasuhiro",
    title = "{Supercurrents in matrix theory and the generalized AdS / CFT correspondence}",
    eprint = "hep-th/0011122",
    archivePrefix = "arXiv",
    reportNumber = "UT-KOMABA-00-14",
    doi = "10.1016/S0550-3213(01)00126-2",
    journal = "Nucl. Phys. B",
    volume = "602",
    pages = "147--171",
    year = "2001"
}

@article{Klebanov:1997kc,
    author = "Klebanov, Igor R.",
    title = "{World volume approach to absorption by nondilatonic branes}",
    eprint = "hep-th/9702076",
    archivePrefix = "arXiv",
    reportNumber = "PUPT-1682",
    doi = "10.1016/S0550-3213(97)00235-6",
    journal = "Nucl. Phys. B",
    volume = "496",
    pages = "231--242",
    year = "1997"
}

@article{Klebanov:1999xv,
    author = "Klebanov, Igor R. and Taylor, Washington and Van Raamsdonk, Mark",
    title = "{Absorption of dilaton partial waves by D3-branes}",
    eprint = "hep-th/9905174",
    archivePrefix = "arXiv",
    reportNumber = "PUPT-1865, MIT-CTP-2866",
    doi = "10.1016/S0550-3213(99)00448-4",
    journal = "Nucl. Phys. B",
    volume = "560",
    pages = "207--229",
    year = "1999"
}

@article{Catterall:2007fp,
    author = "Catterall, Simon and Wiseman, Toby",
    title = "{Towards lattice simulation of the gauge theory duals to black holes and hot strings}",
    eprint = "0706.3518",
    archivePrefix = "arXiv",
    primaryClass = "hep-lat",
    reportNumber = "IMPERIAL-TP-07-TW-03",
    doi = "10.1088/1126-6708/2007/12/104",
    journal = "JHEP",
    volume = "12",
    pages = "104",
    year = "2007"
}

@article{Catterall:2008yz,
    author = "Catterall, Simon and Wiseman, Toby",
    title = "{Black hole thermodynamics from simulations of lattice Yang-Mills theory}",
    eprint = "0803.4273",
    archivePrefix = "arXiv",
    primaryClass = "hep-th",
    doi = "10.1103/PhysRevD.78.041502",
    journal = "Phys. Rev. D",
    volume = "78",
    pages = "041502",
    year = "2008"
}

@article{Catterall:2009xn,
    author = "Catterall, Simon and Wiseman, Toby",
    title = "{Extracting black hole physics from the lattice}",
    eprint = "0909.4947",
    archivePrefix = "arXiv",
    primaryClass = "hep-th",
    doi = "10.1007/JHEP04(2010)077",
    journal = "JHEP",
    volume = "04",
    pages = "077",
    year = "2010"
}

@article{Hanada:2007ti,
    author = "Hanada, Masanori and Nishimura, Jun and Takeuchi, Shingo",
    title = "{Non-lattice simulation for supersymmetric gauge theories in one dimension}",
    eprint = "0706.1647",
    archivePrefix = "arXiv",
    primaryClass = "hep-lat",
    reportNumber = "RIKEN-TH-104, KEK-TH-1158",
    doi = "10.1103/PhysRevLett.99.161602",
    journal = "Phys. Rev. Lett.",
    volume = "99",
    pages = "161602",
    year = "2007"
}

@article{Anagnostopoulos:2007fw,
    author = "Anagnostopoulos, Konstantinos N. and Hanada, Masanori and Nishimura, Jun and Takeuchi, Shingo",
    title = "{Monte Carlo studies of supersymmetric matrix quantum mechanics with sixteen supercharges at finite temperature}",
    eprint = "0707.4454",
    archivePrefix = "arXiv",
    primaryClass = "hep-th",
    reportNumber = "RIKEN-TH-112, KEK-TH-1165",
    doi = "10.1103/PhysRevLett.100.021601",
    journal = "Phys. Rev. Lett.",
    volume = "100",
    pages = "021601",
    year = "2008"
}

@article{Hanada:2008ez,
    author = "Hanada, Masanori and Hyakutake, Yoshifumi and Nishimura, Jun and Takeuchi, Shingo",
    title = "{Higher derivative corrections to black hole thermodynamics from supersymmetric matrix quantum mechanics}",
    eprint = "0811.3102",
    archivePrefix = "arXiv",
    primaryClass = "hep-th",
    doi = "10.1103/PhysRevLett.102.191602",
    journal = "Phys. Rev. Lett.",
    volume = "102",
    pages = "191602",
    year = "2009"
}

@article{Hanada:2008gy,
    author = "Hanada, Masanori and Miwa, Akitsugu and Nishimura, Jun and Takeuchi, Shingo",
    title = "{Schwarzschild radius from Monte Carlo calculation of the Wilson loop in supersymmetric matrix quantum mechanics}",
    eprint = "0811.2081",
    archivePrefix = "arXiv",
    primaryClass = "hep-th",
    doi = "10.1103/PhysRevLett.102.181602",
    journal = "Phys. Rev. Lett.",
    volume = "102",
    pages = "181602",
    year = "2009"
}

@article{Filev:2015hia,
    author = "Filev, Veselin G. and O'Connor, Denjoe",
    title = "{The BFSS model on the lattice}",
    eprint = "1506.01366",
    archivePrefix = "arXiv",
    primaryClass = "hep-th",
    reportNumber = "DIAS-STP-15-09",
    doi = "10.1007/JHEP05(2016)167",
    journal = "JHEP",
    volume = "05",
    pages = "167",
    year = "2016"
}

@article{Wiseman:2008qa,
    author = "Wiseman, Toby and Withers, Benjamin",
    title = "{Holographic renormalization for coincident Dp-branes}",
    eprint = "0807.0755",
    archivePrefix = "arXiv",
    primaryClass = "hep-th",
    doi = "10.1088/1126-6708/2008/10/037",
    journal = "JHEP",
    volume = "10",
    pages = "037",
    year = "2008"
}

@article{Bobev:2025idz,
    author = "Bobev, Nikolay and Mera {\'A}lvarez, Guillermo and Paul, Hynek",
    title = "{Correlation functions for non-conformal Dp-brane holography}",
    eprint = "2503.18770",
    archivePrefix = "arXiv",
    primaryClass = "hep-th",
    doi = "10.1007/JHEP07(2025)137",
    journal = "JHEP",
    volume = "07",
    pages = "137",
    year = "2025"
}
\end{document}